\documentclass[journal]{IEEEtran}

\usepackage[T1]{fontenc}
\usepackage[utf8]{inputenc} % Overleaf 一般可省略，但加上更稳

\usepackage{amsmath,amsfonts}
\usepackage{amssymb}
\usepackage{bm}
\usepackage{bbm}

\usepackage{amsthm}

\usepackage{algorithmic}
\usepackage{algorithm}
\usepackage{array}
\usepackage[caption=false,font=footnotesize,labelfont=rm,textfont=rm]{subfig}
\usepackage{textcomp}
\usepackage{stfloats}
\usepackage{url}
\usepackage{verbatim}
\usepackage{graphicx}
\usepackage{cite}
\usepackage{balance}
\usepackage{xcolor}
\usepackage{multirow}
\usepackage{multicol}
\usepackage{makecell}
\usepackage{tabularx}
\usepackage{booktabs}
\usepackage{titlesec}

\newtheoremstyle{remarkbf}% 〈style name〉
  {3pt}{3pt}               % space above / below
  {\normalfont}            % body font
  {}                       % indent amount
  {\bfseries}              % head font → bold
  {.}{ }{}                 % punctuation, spacing, theorem head spec

\theoremstyle{remarkbf}
\newtheorem{remark}{Remark}

\ifCLASSINFOpdf
\else
\fi
\begin{document}
%
% paper title
% Titles are generally capitalized except for words such as a, an, and, as,
% at, but, by, for, in, nor, of, on, or, the, to and up, which are usually
% not capitalized unless they are the first or last word of the title.
% Linebreaks \\ can be used within to get better formatting as desired.
% Do not put math or special symbols in the title.
\title{Securing the Sensing Functionality in ISAC: KLD-Based Ambiguity Function Shaping}
%
%
% author names and IEEE memberships
% note positions of commas and nonbreaking spaces ( ~ ) LaTeX will not break
% a structure at a ~ so this keeps an author's name from being broken across
% two lines.
% use \thanks{} to gain access to the first footnote area
% a separate \thanks must be used for each paragraph as LaTeX2e's \thanks
% was not built to handle multiple paragraphs
%

\author{Borui~Du,
        Kawon~Han,~\IEEEmembership{Member,~IEEE,}
        and~Christos~Masouros,~\IEEEmembership{Fellow,~IEEE}
        % <-this % stops a space
\thanks{Borui Du and Christos Masouros are with the Department of Electronic and Electrical Engineering, University College London, London WC1E 6BT, U.K. (Corresponding author: Borui Du; email: borui.du.24@ucl.ac.uk, c.masouros@ucl.ac.uk).}% <-this % stops a space
\thanks{Kawon Han is with the Department of Electrical Engineering, Ulsan National Institute of Science and Technology, Ulsan, South Korea (email: kawon.han@unist.ac.kr).}}

% note the % following the last \IEEEmembership and also \thanks - 
% these prevent an unwanted space from occurring between the last author name
% and the end of the author line. i.e., if you had this:
% 
% \author{....lastname \thanks{...} \thanks{...} }
%                     ^------------^------------^----Do not want these spaces!
%
% a space would be appended to the last name and could cause every name on that
% line to be shifted left slightly. This is one of those "LaTeX things". For
% instance, "\textbf{A} \textbf{B}" will typeset as "A B" not "AB". To get
% "AB" then you have to do: "\textbf{A}\textbf{B}"
% \thanks is no different in this regard, so shield the last } of each \thanks
% that ends a line with a % and do not let a space in before the next \thanks.
% Spaces after \IEEEmembership other than the last one are OK (and needed) as
% you are supposed to have spaces between the names. For what it is worth,
% this is a minor point as most people would not even notice if the said evil
% space somehow managed to creep in.

% The paper headers
\markboth{Journal of \LaTeX\ Class Files,~Vol.~14, No.~8, August~2015}%
{Shell \MakeLowercase{\textit{et al.}}: Bare Demo of IEEEtran.cls for IEEE Journals}
% The only time the second header will appear is for the odd numbered pages
% after the title page when using the twoside option.
% 
% *** Note that you probably will NOT want to include the author's ***
% *** name in the headers of peer review papers.                   ***
% You can use \ifCLASSOPTIONpeerreview for conditional compilation here if
% you desire.

% If you want to put a publisher's ID mark on the page you can do it like
% this:
%\IEEEpubid{0000--0000/00\$00.00~\copyright~2015 IEEE}
% Remember, if you use this you must call \IEEEpubidadjcol in the second
% column for its text to clear the IEEEpubid mark.

% use for special paper notices
%\IEEEspecialpapernotice{(Invited Paper)}

% make the title area
\maketitle

% As a general rule, do not put math, special symbols or citations
% in the abstract or keywords.  
\begin{abstract}
As integrated sensing and communication (ISAC) systems are deployed in next-generation wireless networks, a new security vulnerability emerges, particularly in terms of sensing privacy. Unauthorized sensing eavesdroppers (Eve) can potentially exploit the ISAC signal for their own independent passive sensing. However, solutions for sensing-secure ISAC remain largely unexplored to date. This work addresses sensing-security for OFDM- and OTFS-based ISAC waveforms from a target-detection perspective, aiming to prevent Eves from exploiting the ISAC signal for unauthorized passive sensing. We develop ISAC system models for the base station (BS), communication user equipment, and the sensing Eve, and define a Kullback–Leibler-divergence-based detection metric that accounts for mainlobe, sidelobe, and noise components in the ambiguity function and the resulting range-Doppler maps of the legitimate BS's and Eve's sensing. Building on this analysis, we formulate a sensing-secure ISAC signaling design problem that tunes a perturbation matrix to jointly control signal amplitude and phase in the time–frequency domain and solve it via simulated annealing. Simulation results show that the proposed scheme substantially degrades Eve’s detection probability—from 79.4\% to 37.4\% for OTFS and from 94.3\% to 33.0\% for OFDM—while incurring only a small loss in BS sensing performance. In addition, it allows controllable trade-offs across sensing-security and communication performance.
\end{abstract}

% Note that keywords are not normally used for peerreview papers.

% For peer review papers, you can put extra information on the cover
% page as needed:
% \ifCLASSOPTIONpeerreview
% \begin{center} \mathbfseries EDICS Category: 3-BBND \end{center}
% \fi
%
% For peerreview papers, this IEEEtran command inserts a page break and
% creates the second title. It will be ignored for other modes.
\IEEEpeerreviewmaketitle

\section{Introduction}
\label{sec:Intro}
\IEEEPARstart{W}{ith} the rapid expansion of wireless technology, the coverage of wireless devices is increasing significantly, and managing the expanding population of devices and signal types is becoming increasingly costly in terms of spectrum and physical infrastructure. From the perspective of the technology paradigm, the two main functions of radio—communication and sensing—are combinable; therefore, integrated sensing and communication (ISAC) is regarded as a significant direction for future wireless technology development \cite{liu_joint_2020}. Communication uses a radio signal to convey information from the source, through the channel, to the sink, while sensing uses a radio signal to capture information from the source of the channel to the sink—both follow the same transmitter–channel–receiver process. Moreover, to eliminate channel-introduced distortion, communication performs channel estimation and equalization, and channel estimation essentially captures information from the channel as a sensing function, indicating strong potential for combining the two functions in demand and technological essentials.

Given that ISAC uses the same signal to convey information and sense the environment, new security vulnerabilities arise for the ISAC system, which have become a significant concern in recent studies \cite{qu_privacy_2024, chen_survey_2025, su_inte_2025}. This concern is treated as an ISAC security problem with two facets: data security, which protects the confidentiality of communication information during sensing \cite{su_sensing-assisted_2024, liu_secure_2024, chen_survey_2025}, and sensing-security, which protects the sensing targets from unauthorized sensing \cite{qu_privacy_2024, zou_securing_2024,  han_sensing_2025}. Sensing-security risks arise because the ISAC signal is designed for strong communication performance and high sensing quality, which unauthorized users can repurpose for their own sensing, exposing environmental, personal and potentially critical information. Although attention to ISAC security has grown in recent years, most efforts address data security, leaving sensing-security comparatively underdeveloped. Advancing sensing-secured ISAC is therefore increasingly important as the other side of the coin in future ISAC security.

\subsection{Related Works}

Early efforts focused on data security, assuming that a sensing target acts as an eavesdropper (Eve) and attempts to intercept the information-bearing component of the ISAC signal. Su et al.~\cite{su_secure_2021, su_secure_2022} propose two physical-layer data security schemes for multiple-antenna ISAC systems: one minimizes the SNR at Eve while constraining legitimate communication and sensing, and the other exploits destructive interference to deteriorate Eve’s eavesdropping. Afterwards, Su et al.~\cite{su_sensing-assisted_2024} propose a two-stage design that first estimates the target angles and then maximizes the secrecy rate and optimizes the Cramér–Rao bound (CRB) through joint beamforming and artificial-noise (AN) covariance design. Liu et al.~\cite{liu_secure_2024} extend AN-aided protection to orthogonal time frequency space (OTFS) waveforms. They show secrecy gains with or without knowledge of the eavesdropper’s channel state information (CSI) by adapting beamforming or injecting AN into the null space.

Before ISAC sensing-security emerged as a distinct topic, researchers studied sensing privacy in legacy systems, such as Wi-Fi. Ghiro et al.~\cite{ghiro_implementation_2022} and Abanto et al.~\cite{abanto-leon_stay_2020} obfuscate Wi-Fi localization by randomizing preambles or pilots, or by injecting phase noise that corrupts CSI while preserving legitimate sensing. Wang et al.~\cite{wang_multi-antenna_2024} and Hu et al.~\cite{hu_wishield_2024} leverage multiple-input multiple-output (MIMO) capabilities. They use phase-difference injection and neural-network-aided recovery to protect sensing privacy while enabling authorized CSI reconstruction. These methods effectively encrypt CSI. However, because their protection is fundamentally CSI-centric and assumes multi-antenna channel estimation, it does not extend to non-CSI-based or passive sensing that infers objects directly from ambient reflections rather than reconstructed CSI.

Recent work has therefore broadened the security focus from data eavesdropping to sensing-security. Guo et al.~\cite{guo_secure_2024} discuss countermeasures such as cooperative jamming, directional antennas, and beamforming. Qu et al.~\cite{qu_privacy_2024} survey privacy and security in ISAC and formalize attacker models that eavesdrop on motion parameters or jam reflections. In general, the sensing-security solutions can be divided into two types: Eve-aware and Eve-agnostic~\cite{han_2025_next}. The sensing-security problem has been widely investigated in an Eve-aware paradigm, where the BS can leverage some prior information about sensing eavesdroppers. Zou et al.~\cite{zou_securing_2024} propose AN-aided beamforming that protects the sensing function, where a communication user is taken as the sensing eavesdropper, and the design maximizes the mutual information (MI) of legitimate sensing while constraining that of the eavesdropper. Based on an active Eve model, Li et al.~\cite{li_win-win_2025} study an Eve-aware non-orthogonal multiple access-based ISAC system that balances achievable communication rates with CRB-based sensing constraints through joint subcarrier scheduling, beamforming, and jamming.

Note that the above solutions typically build on the availability of Eve’s locations and channels at the legitimate ISAC base station (BS). In some scenarios, however, a sensing Eve may remain silent and thus may not readily reveal their presence, making it difficult to acquire accurate channel knowledge of such eavesdroppers. Consequently, recent research has also explored Eve-agnostic sensing-security in ISAC systems. Wang et al.~\cite{wang_generative_2025} introduce a diffusion-model-based safeguarding signal for Eve-agnostic sensing-security in orthogonal frequency-division multiplexing (OFDM) ISAC systems and report superior performance over existing baselines. Very recently, Han et al.~\cite{han_sensing_2025} explicitly exploit ambiguity function (AF) shaping for OFDM to enhance Eve-agnostic sensing-security by controlling subcarrier power to insert artificial peaks into the range-Doppler (RD) response of Eve’s sensing. Chen et al.~\cite{chen_2025_sensing} design an Eve-agnostic sensing-security scheme for near-field ISAC systems that exploits environmental scatterers to deceive sensing eavesdroppers.

Nevertheless, most existing works remain largely focused on OFDM waveforms and theory-oriented metrics such as MI and CRB for sensing-security design in ISAC systems, and a unified framework with practical sensing-security metrics and multi-waveform support is still unexplored. Other waveform candidates, such as frequency-modulated continuous-wave (FMCW), OTFS, and affine frequency-division multiplexing (AFDM), exhibit distinct sensing–communication trade-offs that call for dedicated sensing-security designs \cite{bonsch_evaluating_2025,hadani_orthogonal_2017,yuan_integrated_2021,bemani_afdm_2021,bemani_integrated_2024}. From an AF perspective, these waveform choices induce different RD structures and sensing-security leakage patterns, providing a unifying lens that links waveform structure to resolvability and unintended sensing. Recent AF-based studies on single-carrier, OFDM, and OTFS clarify how waveform structure governs the sensing–communication trade-off \cite{liu_uncovering_2025,chong_delay-doppler_2025}, while for AFDM, a closed-form AF expression has been derived and parameter regimes with reduced sidelobes have been identified \cite{bedeer_ambiguity_2025}.

\subsection{Motivations \& Contributions}
Current waveform and AF analyses do not adequately account for sensing-security performance, leaving the interaction among waveform design, AF behavior, and sensing-security insufficiently understood. For current ISAC sensing-security studies, two limitations dominate: many studies rely on theoretical metrics and models without validation on practical waveforms, and dependence on artificial noise imposes a power burden that benefits only security. Consequently, ISAC sensing-security remains early-stage, with threat models and evaluation metrics still maturing. In such conditions, this study investigates the sensing-security problem from a detection perspective, aiming to secure the sensing functionality in ISAC systems. The proposed approach is applicable to both Eve-aware and Eve-agnostic cases, thereby bridging the two regimes. Our contributions are summarized as follows:
\begin{itemize}
    \item We develop system models for user equipment (UE) communication, legitimate base station sensing, and Eve’s passive sensing under OFDM and OTFS waveform operations. Building on the system models, we decompose the received sensing RD response at both the BS and Eve into mainlobe, sidelobe, and noise components guided by the Wiener filtering, or so-called linear minimum mean square error (LMMSE) filtering, and matched filtering (MF) receiver properties. This yields closed-form expressions for the effective sensing-signal-to-interference-plus-noise ratio (SINR) contributions for detection under OFDM and OTFS waveform operations. The decomposition links these components to detection performance for legitimate and unauthorized sensing.

    \item We define a Kullback–Leibler divergence (KLD) metric between the sensing distributions for the target-absent and target-present cases at both the BS and Eve. To enable efficient optimization, we apply Jensen’s inequality to the expectation terms and derive a tight, computable surrogate of the derived KLD. This surrogate retains the KLD’s dependence on the waveform design and is therefore suitable as a tractable objective for subsequent sensing-security optimization. Based on the derived KLD, we define the KLD gap between the legitimate BS and Eve as a sensing-security metric.

    \item Using the KLD-based metrics together with a communication-error penalty, we formulate a sensing-security optimization over a perturbation matrix that controls amplitude and phase in the time-frequency (TF) domain under standard power constraints. In this optimization, Eve's KLD is expressed in terms of Eve's channel coefficients and noise power allowing for an Eve-aware design. In practice, however, these quantities can be set to arbitrary values rather than Eve's exact parameters, so that the method can be used in Eve-agnostic scenarios. We solve the resulting nonconvex problem using simulated annealing.

\end{itemize}

Our Monte Carlo simulations show significant gains: the proposed design consistently elevates Eve’s sidelobes and reduces their detection probability while keeping the BS’s sensing degradation small. By varying the priority weight between sensing-security and communication, the proposed method demonstrates security improvements and characterizes the sensing-security–communication trade-off. It reveals that stronger sensing-security introduces a graceful degradation of communication errors and the data rate, while lowering Eve’s detection probability, thereby enhancing resilience to unauthorized sensing.

The rest of this paper is organized as follows. Section \ref{sec:Models} presents the system model, including the scenario model and its assumptions, followed by the ISAC signal model for BS, UE, and Eve. Section \ref{sec:sen_analy} presents the mainlobe, sidelobe, and noise component decomposition of the RD response and a KLD-based secure sensing analysis for both BS and Eve. The optimization problem is presented in Section \ref{sec:opt_prob} and is solved using a simulated annealing algorithm due to its complex structure. The simulation results and interpretation are presented in Section \ref{sec:Num_res}. Finally, the conclusion is formulated in Section \ref{sec:conclu}.

In this paper, matrices and vectors are denoted by bold letters, e.g., \( \mathbf{A} \) and \( \mathbf{a} \) or \(\underline{a}\), while \(\left[\mathbf{A}\right]_{i,j}\) represents the element in the \(i\)-th row and \(j\)-th column of matrix \(\mathbf{A}\). \( \mathbb{C} \) and \( \mathbb{R} \) represent the complex and real number sets, respectively. The operators \( (\cdot)^T \), \( (\cdot)^* \), and \( (\cdot)^H \) denote matrix transpose, conjugate, and conjugate transpose, respectively. The notation \( \mathcal{CN}(0, \mathbf{C}) \) represents a zero-mean complex Gaussian distribution with covariance matrix \( \mathbf{C} \). $\odot$ and $\oslash$ are element-wise product and division. The trace, vectorization, and Kronecker product operators are denoted as \( \operatorname{Tr}(\cdot) \), \( \operatorname{vec}(\cdot) \), and \( \otimes \), respectively. \( \operatorname{vec}^{-1}(\cdot) \) is the reshape function that transforms a vector into a matrix, and $\operatorname{diag}\left(\cdot\right)$ is the diagonal matrix operator. Expectation is represented by \( \mathbb{E}\{\cdot\} \). The normalized discrete Fourier matrix and the identity matrix of dimension \( N \) are denoted as \( \mathbf{F}_N \) and \( \mathbf{I}_N \), respectively. $|\cdot|$, $\|\cdot\|_2$ and $\|\cdot\|_F$ are element-wise absolute value, $l_2$ norm, and Frobenius norm.

\section{System Model}
\label{sec:Models}
\begin{figure}
    \centering
    \includegraphics[width=0.95\linewidth]{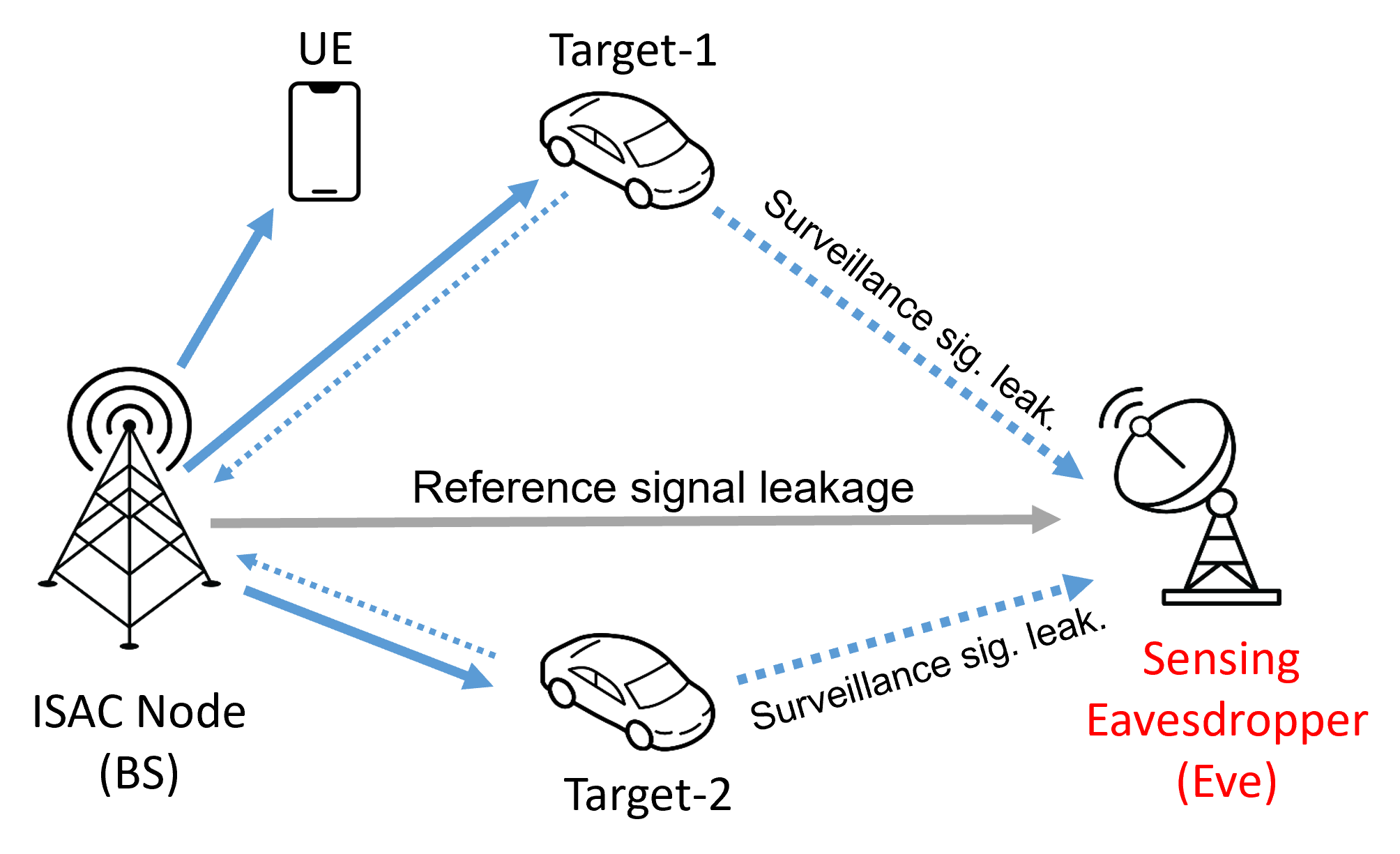}
    \caption{The illustration of the scenario model.}
    \label{fig:scenario_mod}
\end{figure}

In this section, we first specify the system model and its underlying assumptions. The ISAC system executes joint sensing and communications via a single BS. We focus on a specific scenario that contains one UE, multiple targets, and one passive sensing eavesdropper. The BS radiates a unified ISAC waveform through a single-input single-output antenna that simultaneously: (i) delivers data to the UE and (ii) probes the environment to detect the existence of targets, in a given range-Doppler bin. The Eve, while silent, attempts to detect the same targets in a passive way, by utilizing the BS’s waveform. The illustration of the scenario model is presented in Fig.~\ref{fig:scenario_mod}.

To facilitate the analysis and focus on the essential property of the sensing-security scenario, we impose some assumptions for the system \cite{han_sensing_2025}. Particularly, we develop a signaling design that is applicable to both Eve-aware and Eve-agnostic scenarios.
\begin{itemize}
    \item \textbf{A.1:} Depending on the scenario, the BS may have or lack knowledge of Eve-related parameters (e.g., target-channel coefficients and noise power), covering both Eve-aware and Eve-agnostic settings.    
    \item \textbf{A.2:} Eve knows the BS’s location, synchronization parameters, and carrier frequency, but is unaware of the exact transmitted signal structure, where the transmitted signal carries random data and no pilot symbols.
    \item \textbf{A.3:} Eve has no prior information about the target.
    \item \textbf{A.4:} Eve possesses sufficient spatial resources (e.g., antenna elements) to extract a reference signal with enough separation from their direct channel to the BS.  
    \item \textbf{A.5:} The target moves slowly compared to the resolution of the transmitted signal. Therefore, the sensing function can be performed using TF domain processing with the approximation in \cite{raviteja_orthogonal_2019}.
\end{itemize}

\subsection{Transmitted Signal Model}

The transmitted OFDM and OTFS signals can be represented as a general form in the TF domain with perturbation, which is expressed as
\begin{equation}
    \mathbf{x}_\text{TF} = \mathbf{W}_\text{TF}\mathbf{s}_\text{TF} = \mathbf{W}_\text{TF}\mathbf{U}\mathbf{s}\label{equ:sig_mod},
\end{equation}
where $\mathbf{s}_{{\text{TF}}} \in \mathbb{C}^{MN \times 1}$ is the data symbol in the TF domain, and $\mathbf{U} \in \mathbb{C}^{MN \times MN}$ is the transform matrix for different modulation methods. $M$ is the number of subcarriers, and $N$ is the number of time-slots. For OFDM, $\mathbf{U} = \mathbf{I}_{MN}$, indicating that the communication symbol coincides with the TF samples. For OTFS, $\mathbf{U} = \left(\mathbf{F}_{N}^{H} \otimes \mathbf{F}_{M}\right)$, which represents the inverse symplectic finite Fourier transform to map the delay-Doppler (DD) domain symbols into the TF domain. Note that $\mathbf{U}$ is a unitary matrix for both OFDM and OTFS, and typically for other waveforms. Throughout this paper, we model the data vector $\mathbf{s}\in\mathbb{C}^{MN}$ as having independent, identically distributed (i.i.d.) elements with zero mean and unit variance but with specific constellations, i.e., $\mathbb{E}\{\mathbf{s}\}= \mathbf{0}$ and $\mathbb{E}\{\mathbf{s}\mathbf{s}^H\}=\mathbf{I}_{MN}$. \noindent $\mathbf{W}_{\text{TF}} \in \mathbb{C}^{MN \times MN}$ denotes the complex perturbation matrix in the TF domain, enabling both amplitude and phase control; specifically, $\mathbf{W}_{\text{TF}}=\operatorname{diag}([w_1,\ldots,w_{MN}]^{T})$, where each $w_i\in\mathbb{C}$ applies an independent per TF-bin perturbation, assuming
\begin{align}
    \operatorname{Tr}\left({{\mathbf{W}}^H_{{\text{TF}}}}{{\mathbf{W}}_{{\text{TF}}}}\right) = {\sum\limits^{MN}_{i=1}}{|w_{i}|^2} = MN.
\end{align}
Consequently, the total energy of the modulated signal is $MN$, while the power for each symbol is controlled by $\mathbf{W}_{\text{TF}}$. Then, the TF domain signal, $\mathbf{x}_\text{TF}$, is transformed using inverse fast Fourier transform for transmission, which can be expressed as follows
\begin{align}
    \mathbf{x}_\text{t} = \left(\mathbf{I}_{N}\otimes\mathbf{F}^H_{M}\right) \mathbf{x}_\text{TF},
\end{align}
where $\mathbf{x}_\text{t}\in\mathbb{C}^{MN \times 1}$ is the transmitted signal in the time domain.

\subsection{Communication Signal Model}
Using the Rician model, the communication channel between the BS and UE can be expressed as follows
\begin{align}
    \mathbf{H}_\text{c} = \alpha_c\sqrt{\frac{\kappa_c}{\kappa_c + 1}}\, \mathbf{\Pi}_{MN}^{l_c}\mathbf{\Delta}_{MN}^{k_c} + \alpha_c\sqrt{\frac{1}{\kappa_c + 1}}\mathbf{H}_\text{c,NLoS},
\end{align}
where $\mathbf{H}_\text{c} \in \mathbb{C}^{MN \times MN}$ and $\alpha_c, \kappa_c$, $l_c$, and $k_c$ are the path-loss, Rician factor, delay, and Doppler shift of the communication channel, respectively\footnote{In this work, we assume integer delay and Doppler for brevity, and $l^\text{BS}_{s,p}$, $k^\text{BS}_{s,p}$, $l^\text{E}_{s,p}$, and $k^\text{E}_{s,p}$ are normalized by the corresponding resolutions in the simulation section \cite{raviteja_practical_2019}.}. The addition and removal of the cyclic-prefix (CP) are assumed to be performed before parameter estimation. Therefore, the delay operation in the delay domain is modeled as a circular shift, i.e., $\mathbf{\Pi}_{MN}^{l_c}$ and $\mathbf{\Delta}_{MN}^{k_c}$ are delay shift and Doppler phase matrices as shown in \cite{raviteja_practical_2019}. Besides, the elements of $\mathbf{H}_\text{c,NLoS}$ follow $\mathcal{CN}\left(0, 1\right)$, which represents the non-line-of-sight (NLoS) channel in the transmission.

The received signal in the symbol multiplexing domain for the UE can be formulated as
\begin{subequations}
    \begin{align}
    \mathbf{y}_\text{sym} &= \underbrace{\mathbf{U}^H\left(\mathbf{I}_{N}\otimes\mathbf{F}_{M}\right)\mathbf{H}_\text{c}\left(\mathbf{I}_{N}\otimes\mathbf{F}^H_{M}\right)}_{\triangleq\mathbf{H}_\text{c,eff}}\mathbf{W}_\text{TF}\mathbf{U}\mathbf{s} \notag \\ 
    & \quad+ \underbrace{\mathbf{U}^H\left(\mathbf{I}_{N}\otimes\mathbf{F}_{M}\right)\mathbf{z}_\text{c}}_{\triangleq\mathbf{z}_\text{c,eff}} \\
    &= \mathbf{H}_\text{c,eff}\mathbf{W}_\text{TF}\mathbf{U}\mathbf{s} + \mathbf{z}_\text{c,eff},
    \end{align}
\end{subequations}
where $\mathbf{z}_\text{c,eff} \sim \mathcal{CN}\left(\mathbf{0}, \sigma^2_{\mathbf{z}_\text{c}}\mathbf{I}_{MN}\right)$ is the effective additive white Gaussian noise (AWGN) after the inverse unitary transform of $\mathbf{U}^H\left(\mathbf{I}_{N}\otimes\mathbf{F}_{M}\right)$, and the $\sigma^2_{\mathbf{z}_\text{c}}$ is the variance of the noise for the communication function. Therefore, the achievable rate of a specific transmission is
\begin{align}
    {R_{\text{c}}} = \log \left[ {1 + \frac{{{{\left\| {\mathbf{s}} \right\|}^2_2}}}{{{{\left\| {{{\mathbf{H}}_{{\text{c}},{\text{eff}}}}{{\mathbf{W}}_{{\text{TF}}}}{\mathbf{Us}} - {\mathbf{s}}} \right\|}^2_2} + {{\left\| {{{\mathbf{z}}_{{\text{c}},{\text{eff}}}}} \right\|}^2_2}}}} \right],
\end{align}
where ${{{\left\| {{{\mathbf{H}}_{{\text{c}},{\text{eff}}}}{{\mathbf{W}}_{{\text{TF}}}}{\mathbf{Us}} - {\mathbf{s}}} \right\|}^2_2}}$ is the mismatch component of the received signal at the UE's side \cite{zhang_dual_2024, song_low_2025}. Consequently, minimizing the mismatch component based on the channel information of ${{\mathbf{H}}_{{\text{c}},{\text{eff}}}}$ is equivalent to maximizing the achievable rate. Therefore, this mismatch component serves as the communication metric for the sensing-secured ISAC design \cite{zhang_dual_2024, song_low_2025}.

\subsection{Legitimate Monostatic Sensing Signal Model}

The BS performs sensing in a monostatic mode. Based on the line-of-sight (LoS) channel model and assuming the velocity is small (\textbf{A.5}), we have the following expression of the sensing channel representation in the TF domain \
\begin{subequations}
    \begin{align}
        \mathbf{H}_\text{BS,s} &= \sum^{K}_{p=1}{\alpha^\text{BS}_{s,p}} \cdot \underline{\psi}^H\left({l^\text{BS}_{s,p}}\right)\underline{\phi}\left(k^\text{BS}_{s,p}\right),\\
        \text{with } \underline{\psi}\left({l^\text{BS}_{s,p}}\right) &= \operatorname{exp}\left(\left[1, j2\pi \frac{l^\text{BS}_{s,p}}{M}, \cdots,j2\pi \frac{l^\text{BS}_{s,p}(M-1)}{M}\right]\right),\\
        \text{and } \underline{\phi}\left({k^\text{BS}_{s,p}}\right) &= \operatorname{exp}\left(\left[1, j2\pi \frac{k^\text{BS}_{s,p}}{N}, \cdots,j2\pi \frac{k^\text{BS}_{s,p}(N-1)}{N}\right]\right),
    \end{align}
\end{subequations}
where $\mathbf{H}_\text{BS,s} \in \mathbb{C}^{M\times N}$ and $K$ is the number of targets in the scenario. The path-loss coefficients \(\alpha^\text{BS}_{s,p}\) correspond to the random path-loss for target-$p$, including radar cross section, which follows $\alpha^\text{BS}_{s,p}\sim \mathcal{CN}\left(0, \left|\alpha^\text{BS}_{s,p,0}\right|^2\right)$. Additionally, $l^\text{BS}_{s,p}$ and $k^\text{BS}_{s,p}$ are round-trip delay and Doppler shifts, respectively. \(\underline{\psi}\) and \(\underline{\phi}\) are the delay and Doppler phase vectors. Thus, the received sensing signal of BS can be written as
\begin{align}
    \mathbf{R}_\text{BS,s} &= \mathbf{H}_\text{BS,s}\odot\mathbf{X}_\text{TF} + \mathbf{Z}_\text{BS,s,TF}, \label{equ:sensing_rec_signal_BS}
\end{align}
where \(\mathbf{R}_\text{BS,s} \in \mathbb{C}^{M \times N}\), and $\mathbf{X}_\text{TF} \triangleq \operatorname{vec}^{-1}\left(\mathbf{x}_\text{TF}\right)$ represents the samples in the TF domain. $\mathbf{Z}_\text{BS,s,TF}\triangleq\operatorname{vec}^{-1}\left[\left(\mathbf{I}_{N}\otimes\mathbf{F}_{M}\right)\mathbf{z}_\text{BS,s}\right]$ is the noise transformed into the TF domain. $\mathbf{z}_\text{BS,s} \sim \mathcal{CN}\left(0, \sigma^2_{\mathbf{z}_\text{BS,s}}\mathbf{I}_{MN}\right)$, and $\sigma^2_{\mathbf{z}_\text{BS,s}}$ is defined as the noise variance for BS's sensing functionality.

\subsection{Eve Sensing Signal Model}

The Eve performs passive sensing in a bistatic mode. Similar to the LoS sensing channel of the BS, the sensing target for Eve can be represented as follows
\begin{align}
    \mathbf{H}_\text{E,s} &= \sum^{K}_{p=1}{\alpha^\text{E}_{s,p}} \cdot \underline{\psi}^H\left({l^\text{E}_{s,p}}\right)\underline{\phi}\left(k^\text{E}_{s,p}\right),
\end{align}
where $\mathbf{H}_\text{E,s} \in \mathbb{C}^{M \times N}$ and $\alpha^\text{E}_{s,p}$, $l^\text{E}_{s,p}$ and $k^\text{E}_{s,p}$ are different from 
$\alpha^{\text{BS}}_{s,p}$, $l^\text{BS}_{s,p}$, and $k^\text{BS}_{s,p}$ given the different geometry between BS's and Eve's perspectives. Additionally, $\alpha^{\text{E}}_{s,p}\sim\mathcal{CN}\left(0, \left|\alpha^{\text{E}}_{s,p,0}\right|^2\right)$. The received signal model of Eve can be represented as
\begin{align}
    \mathbf{R}_\text{E,s} = \mathbf{H}_\text{E,s}\odot \mathbf{X}_\text{TF} + \mathbf{Z}_\text{E,s,TF},
\end{align}
where $\mathbf{R}_\text{E,s} \in \mathbb{C}^{M \times N}$ and $\mathbf{Z}_\text{E,s,TF}$ is the noise in the TF domain for Eve. $\mathbf{z}_\text{E,s} \sim \mathcal{CN}\left(0, \sigma^2_{\mathbf{z}_\text{E,s}}\mathbf{I}_{MN}\right)$, and $\sigma^2_{\mathbf{z}_\text{E,s}}$ is defined as the noise variance for Eve's sensing functionality. 

Because Eve does not know the transmitted signal from the BS, they need to extract the reference signal from the direct channel, which is assumed to be a Rician channel and can be represented as
\begin{align}
    \mathbf{H}_\text{E,d} &= \alpha^\text{E}_d\sqrt{\frac{\kappa_e}{\kappa_e + 1}}\mathbf{H}_\text{E,d,LoS} + \alpha^\text{E}_d\sqrt{\frac{1}{\kappa_e + 1}}\mathbf{H}_\text{E,d,NLoS},
\end{align}
where $\mathbf{H}_\text{E,d}\in \mathbb{C}^{M \times N}$ and $\alpha^\text{E}_d$ is the path-loss of the direct channel between BS and Eve, and $\kappa_e$ is the Rician factor, and 
\begin{align}
    \mathbf{H}_\text{E,d,LoS} &= \underline{\psi}^H\left({l^\text{E}_d}\right)\underline{\phi}\left(k^\text{E}_d\right),
\end{align}
where $\mathbf{H}_\text{E,d,LoS}\in \mathbb{C}^{M \times N}$ and $l^\text{E}_{d}$ and $k^\text{E}_{d}$ are the motion parameters for the direct channel of Eve. The elements of $\mathbf{H}_\text{E,d,NLoS}$ follow $\mathcal{CN}\left(0, 1\right)$. Therefore, the received signal from the direct channel has the following form
\begin{align}
    \mathbf{R}_\text{E,d} &= \mathbf{H}_\text{E,d}\odot\mathbf{X}_\text{TF} + \mathbf{Z}_\text{E,d,TF} \label{equ:sensing_rec_signal_E},
\end{align}
where $\mathbf{R}_\text{E,d} \in \mathbb{C}^{M \times N}$ and $\mathbf{Z}_\text{E,d,TF}$ is the noise in the TF domain for Eve's direct link. $\mathbf{z}_\text{E,d} \sim \mathcal{CN}\left(0, \sigma^2_{\mathbf{z}_\text{E,d}}\mathbf{I}_{MN}\right)$, and $\sigma^2_{\mathbf{z}_\text{E,d}}$ is defined as the noise variance for Eve's direct link towards the BS.

Given that Eve can infer the LoS component of the direct channel, $\mathbf{H}_\text{E,d,LoS}$, this component can be canceled on Eve's side. The reference signal in the TF domain can be represented as follows
\begin{align}
    \widehat{\mathbf{X}}_\text{E,d,TF}&=\alpha^\text{E}_d\sqrt{\frac{\kappa_e}{\kappa_e + 1}}\mathbf{X}_\text{TF} + \alpha^\text{E}_d\sqrt{\frac{1}{\kappa_e + 1}}\widetilde{\mathbf{H}}_\text{E,d,TF} \odot \mathbf{X}_\text{TF} \notag\\
    &\quad + \widetilde{\mathbf{Z}}_\text{E,d,TF}, \label{equ:sensing_ref_signal_BS}
\end{align}
where $\widetilde{\mathbf{H}}_\text{E,d,TF} \triangleq \mathbf{H}_\text{E,d,NLoS} \oslash \mathbf{H}_\text{E,d,LoS}$ and $\widetilde{\mathbf{Z}}_\text{E,d,TF} \triangleq  \mathbf{Z}_\text{E,d,TF} \oslash \mathbf{H}_\text{E,d,LoS}$. Consequently, (\ref{equ:sensing_ref_signal_BS}) serves as the reciprocal filter (RF) to recover the reference signal. Subsequently, Eve employs $\widehat{\mathbf{X}}_\text{E,d,TF}$ as the reference and applies a receive filter for passive sensing.

\section{ISAC Sensing-Security Analysis}
\label{sec:sen_analy}

This section presents the sensing receiver processing for the BS and Eve, then decomposes the contributions of the mainlobe, sidelobes, and noise on their respective sensing performance. Building on this analysis, we develop a KLD-based analysis of detection performance at the BS and Eve, defining the sensing-security metric as the KLD gap between the legitimate BS and Eve for secure-sensing signaling design.

\subsection{BS Sensing Receiver Processing}
Since the legitimate BS has full knowledge of the transmitted signal $\mathbf{X}_\text{TF}$, the target parameters can be estimated using an LMMSE filter. The LMMSE filter is known to outperform the RF and MF by controlling sidelobes given the signal and noise statistics \cite{Furkan_2025_Fun}, as follows
\begin{align}
    \left[\mathbf{G}_\text{BS}\right]_{m,n} = \frac{x_{m,n}^*}{\left|x_{m,n}\right|^2 + \sigma^2_{\mathbf{z}_\text{BS,s}}} = \frac{w_{m,n}^* s_{m,n}^*}{ \left|w_{m,n}s_{m,n}\right|^2 + \sigma^2_{\mathbf{z}_\text{BS,s}}},
\end{align}
where $x_{m,n} = \left[\mathbf{X}_\text{TF}\right]_{m,n} = w_{m,n} s_{m,n}$. $w_{m,n} \triangleq w_{nM + m}$, $s_{m,n} \triangleq \left[\mathbf{S}_\text{TF}\right]_{m,n}$, and $\mathbf{S}_\text{TF} \in \mathbb{C}^{M\times N} \triangleq \operatorname{vec}^{-1}\left(\mathbf{s}_\text{TF}\right)$.
The filtered signal in the TF domain has the following form
\begin{subequations}
\begin{align}
    \left[\widehat{\mathbf{H}}_\text{BS,s}\right]_{m,n} &= \left[\mathbf{G}_\text{BS}\right]_{m,n} \cdot \left[\mathbf{R}_\text{BS,s}\right]_{m,n}\\
    &= \frac{\left|x_{m,n}\right|^2}{\left|x_{m,n}\right|^2 + \sigma^2_{\mathbf{z}_\text{BS,s}}} \left[\mathbf{H}_\text{BS,s}\right]_{m,n} \notag\\ 
    &\quad+ \frac{ x_{m,n}^*}{\left|x_{m,n}\right|^2 + \sigma^2_{\mathbf{z}_\text{BS,s}}}\left[\mathbf{Z}_\text{BS,s,TF}\right]_{m,n},
\end{align}
\end{subequations}
where $\widehat{\mathbf{H}}_\text{BS,s} \in \mathbb{C}^{M\times N}$ is the estimated channel matrix in the TF domain. Using the symplectic finite Fourier transform, the estimated channel will be transformed into the DD domain, i.e., RD map, as follows
\begin{align}
    \widehat{\mathbf{\mathbf{\Lambda}}}_\text{BS} = \mathbf{F}^H_{M}\widehat{\mathbf{H}}_\text{BS,s}\mathbf{F}_{N},
\end{align}
which can be detailed as
\begin{align}
    \bigl[\widehat{\mathbf{\mathbf{\Lambda}}}_{\text{BS}}\bigr]_{l,k}
    &= \frac{1}{\sqrt{MN}}
    \sum_{m,n}
    \bigl[\widehat{\mathbf{H}}_{\text{BS,s}}\bigr]_{m,n}
    e^{\,j2\pi \left(\frac{m\,l}{M} - \frac{n\,k}{N}\right)}, \label{equ:channel_decomp} % \\
    % &= \frac{1}{\sqrt{MN}}
    % \sum_{m=0}^{M-1}
    % \sum_{n=0}^{N-1}
    % \frac{\left|x_{m,n}\right|^2}{\left|x_{m,n}\right|^2 + \sigma^2_{\mathbf{z}_\text{BS,s}}} \left[\mathbf{H}_\text{BS,s}\right]_{m,n}
    % e^{\,j2\pi \left(\frac{m\,l}{M} - \frac{n\,k}{N}\right)} \\
    % &+ \frac{1}{\sqrt{MN}}
    % \sum_{m=0}^{M-1}
    % \sum_{n=0}^{N-1}
    % \frac{ x_{m,n}^*}{\left|x_{m,n}\right|^2 + \sigma^2_{\mathbf{z}_\text{BS,s}}}\left[\mathbf{Z}_\text{BS,s,TF}\right]_{m,n}
    % e^{\,j2\pi \left(\frac{m\,l}{M} - \frac{n\,k}{N}\right)}
\end{align}
where $l = 0, 1, \cdots, M - 1$ and $k = 0, 1, \cdots, N - 1$. Therefore, the estimate $\widehat{\mathbf{\mathbf{\Lambda}}}_{\text{BS}}$ can be decomposed into target channel component and noise components as below
\begin{subequations}
\begin{align}
    \bigl[\widehat{\mathbf{\mathbf{\Lambda}}}_{\text{BS}}\bigr]_{l,k}
    &=\frac{1}{\sqrt{MN}}
    \sum_{m,n}\frac{\left|x_{m,n}\right|^2}{\left|x_{m,n}\right|^2 + \sigma^2_{\mathbf{z}_\text{BS,s}}} \left[\mathbf{H}_\text{BS,s}\right]_{m,n} \notag\\ 
    &\quad\times\exp\left\{\,j2\pi \left(\tfrac{m\,l}{M} - \tfrac{n\,k}{N}\right)\right\} \notag\\
    &\quad+ \frac{1}{\sqrt{MN}}\sum_{m,n}\frac{ x_{m,n}^*}{\left|x_{m,n}\right|^2 + \sigma^2_{\mathbf{z}_\text{BS,s}}}\left[\mathbf{Z}_\text{BS,s,TF}\right]_{m,n}\notag\\
    &\quad\times\exp\left\{\,j2\pi \left(\tfrac{m\,l}{M} - \tfrac{n\,k}{N}\right)\right\}\\
    &\triangleq \bigl[\widehat{\mathbf{\mathbf{\Lambda}}}^{\text{BS}}_{\mathbf{H}}\bigr]_{l,k} + \bigl[\widehat{\mathbf{\mathbf{\Lambda}}}^{\text{BS}}_{\mathbf{Z}}\bigr]_{l,k}.
\end{align}
\end{subequations}
Given the target-$p$ is located at $(l^\text{BS}_{s,p}, k^\text{BS}_{s,p})$, the channel component, $\bigl[\widehat{\mathbf{\mathbf{\Lambda}}}^{\text{BS}}_{\mathbf{H}}\bigr]_{l^{\text{BS}}_{s,p},k^{\text{BS}}_{s,p}}$, of the estimate RD map at $(l^\text{BS}_{s,p}, k^\text{BS}_{s,p})$ can be simplified as
\begin{subequations}
\begin{align}
\bigl[\widehat{\mathbf{\mathbf{\Lambda}}}^{\text{BS}}_{\mathbf{H}}\bigr]_{l^{\text{BS}}_{s,p},k^{\text{BS}}_{s,p}}
 &= \frac{1}{MN}
    \sum_{m,n}
    \frac{\left|x_{m,n}\right|^2}{\left|x_{m,n}\right|^2 + \sigma^2_{\mathbf{z}_\text{BS,s}}}
    \sum_{l,k}
    [\mathbf{\Lambda}_{\text{BS,s}}]_{l,k} \notag \\
&\quad\times \exp\!\Bigl\{j2\pi\!\bigl(
        \tfrac{m\,(l^{\text{BS}}_{s,p}-l)}{M}
       +\tfrac{n\,(k-k^{\text{BS}}_{s,p})}{N}
    \bigr)\Bigr\},\label{eq:second}\\[4pt]
&= \frac{1}{MN}
    \sum_{m,n}
    \frac{|x_{m,n}|^{2}}{|x_{m,n}|^{2}+\sigma^{2}_{\mathbf{z}_{\text{BS,s}}}}
    {\alpha^{\text{BS}}_{s,p}} \notag\\
&\quad+ \frac{1}{MN}
    \sum_{\substack{q=1\\q\neq p}}^{K}
    \sum_{m,n}
    \frac{|x_{m,n}|^{2}}{|x_{m,n}|^{2}+\sigma^{2}_{\mathbf{z}_{\text{BS,s}}}}
    {\alpha^{\text{BS}}_{s,q}}\notag\\
 &\quad\times
    \exp\!\Bigl\{j2\pi\!\bigl(
        \tfrac{m\,(l^{\text{BS}}_{s,p}-l^{\text{BS}}_{s,q})}{M}
       +\tfrac{n\,(k^{\text{BS}}_{s,q}-k^{\text{BS}}_{s,p})}{N}
    \bigr)\Bigr\},
    \label{eq:Lambda_H_BS}
\end{align}
\end{subequations}
where $\mathbf{\Lambda}_\text{BS,s}$ is the actual target in the RD map from the BS's perspective. We further define the mainlobe and sidelobe interference component from other targets in (\ref{eq:Lambda_H_BS}) as follows
\begin{align}
    \bigl[\widehat{\mathbf{\mathbf{\Lambda}}}^{\text{BS}}_{\mathbf{S}}\bigr]_{l^{\text{BS}}_{s,p},k^{\text{BS}}_{s,p}}\triangleq&\frac{1}{MN}
    \sum_{m,n}
    \frac{|x_{m,n}|^{2}}{|x_{m,n}|^{2}+\sigma^{2}_{\mathbf{z}_{\text{BS,s}}}}
    {\alpha^{\text{BS}}_{s,p}}, \\
    \bigl[\widehat{\mathbf{\mathbf{\Lambda}}}^{\text{BS}}_{\mathbf{I}}\bigr]_{l^{\text{BS}}_{s,p},k^{\text{BS}}_{s,p}} \triangleq&\frac{1}{\sqrt{MN}}
    \sum_{\substack{q=1\\q\neq p}}^{K}{\alpha^{\text{BS}}_{s,q}}\bigl[\widetilde{\mathbf{\mathbf{\Lambda}}}_{\text{BS}}\bigr]_{(l^\text{BS}_{s,p}-l^\text{BS}_{s,q}),(k^\text{BS}_{s,p}-k^\text{BS}_{s,q})},
\end{align}
where $\widetilde{\mathbf{\mathbf{\Lambda}}}_{\text{BS}}$ is the RD response of the LMMSE filter for the detection at BS's side when the target is located at $l^\text{BS}_{s} = 0$ and $k^\text{BS}_{s} = 0$. Therefore, 
\begin{align}
    \bigl[\widehat{\mathbf{\mathbf{\Lambda}}}^{\text{BS}}_{\mathbf{H}}\bigr]_{l^{\text{BS}}_{s,p},k^{\text{BS}}_{s,p}} = \bigl[\widehat{\mathbf{\mathbf{\Lambda}}}^{\text{BS}}_{\mathbf{S}}\bigr]_{l^{\text{BS}}_{s,p},k^{\text{BS}}_{s,p}} + \bigl[\widehat{\mathbf{\mathbf{\Lambda}}}^{\text{BS}}_{\mathbf{I}}\bigr]_{l^{\text{BS}}_{s,p},k^{\text{BS}}_{s,p}}.
\end{align}
Besides, the noise component can be written as
\begin{align}
    \bigl[\widehat{\mathbf{\mathbf{\Lambda}}}^{\text{BS}}_{\mathbf{Z}}\bigr]_{l,k}
    &= \frac{1}{\sqrt{MN}}
    \sum_{m,n}
    \frac{ x_{m,n}^*}{\left|x_{m,n}\right|^2 + \sigma^2_{\mathbf{z}_\text{BS,s}}}\bigl[\mathbf{Z}_\text{BS,s,TF}\bigr]_{m,n} \notag\\
   &\quad\times\exp\left[{\,j2\pi\!\Bigl(\tfrac{m\,l}{M}
                 -\tfrac{n\,k}{N}\Bigr)}\right],
\end{align}
where $l = 0, 1, \cdots, M - 1$ and $k = 0, 1, \cdots, N - 1$. Therefore, the mainlobe component can be applied to detect and estimate the parameters of target-$p$.

Given the existence of a target at $l_t$ and $k_t$ range and Doppler bin, the detection SINR can be defined as follows
\begin{align}
    \text{SINR} = \frac{{{{\mathbb{E} \left\{\left| { {\bigl[\widehat{\mathbf{\mathbf{\Lambda}}}_{\mathbf{S}}\bigr]_{l_{t},k_{t}}} } \right|^2\right\}}}}}{{{ { \mathbb{E} \left\{\left|{\bigl[\widehat{\mathbf{\mathbf{\Lambda}}}_{\mathbf{I}}\bigr]_{l_{t},k_{t}}}\right|^2 \right\}} }} + {\mathbb{E} \left\{ \left| {{\bigl[\widehat{\mathbf{\mathbf{\Lambda}}}_{\mathbf{Z}}\bigr]_{l_{t},k_{t}}}} \right|^2 \right\}}}.
    \label{equ:SINR}
\end{align}
To make this ratio explicit, we next detail the signal and noise powers for the BS and Eve.

Given a specific sequence of data transmission, the numerator, i.e., the expected norm-squared mainlobe for the estimated RD map, has the following expression for the BS:
\begin{subequations}
\begin{align}
    P_{s,p,\text{BS}}
    &\triangleq {{\mathbb{E}\left\{{\left| {{{\bigl[\widehat{\mathbf{\mathbf{\Lambda}}}^{\text{BS}}_{\mathbf{S}}\bigr]_{l^\text{BS}_{s,p},k^\text{BS}_{s,p}}}}} \right|}^2\right\}}} \\
    &= \frac{\left|\alpha^{\text{BS}}_{s,p,0}\right|^2}{\left(MN\right)^2}
   \left|\sum_{m,n}
   \frac{\lvert x_{m,n}\rvert^2}{\lvert x_{m,n}\rvert^2 + \sigma^2_{\mathbf{z}_\text{BS,s}}}\right|^2,
\end{align}
\end{subequations}
where $x_{m,n}\triangleq \left[\mathbf{X}_\text{TF}\right]_{m,n}$, and 
\begin{align}
    \mu_{p,\text{BS}}\triangleq\frac{\alpha^{\text{BS}}_{s,p,0}}{MN}\sum_{m,n}\frac{\lvert x_{m,n}\rvert^2}{\lvert x_{m,n}\rvert^2 + \sigma^2_{\mathbf{z}_\text{BS,s}}}.
\end{align}
Having established the signal terms, we now quantify the associated noise contributions. The noise power for the BS is
\begin{subequations}
\begin{align}
    P_{n,\text{BS}}&\triangleq{\mathbb{E} \left\{ \left| {{\bigl[\widehat{\mathbf{\mathbf{\Lambda}}}^{\text{BS}}_{\mathbf{Z}}\bigr]_{l^\text{BS}_{s,p},k^\text{BS}_{s,p}}}} \right|^2 \right\}} \\
    &= \frac{\sigma^2_{\mathbf{z}_\text{BS,s}}}{MN}\sum_{m,n}
    \frac{\lvert x_{m,n}\rvert^2}{\left(\lvert x_{m,n}\rvert^2 + \sigma^2_{\mathbf{z}_\text{BS,s}}\right)^2}.
\end{align}
\end{subequations}
The interference produced by other targets is expressed as follows, assuming sufficient independence of each target, thereby rendering the cross-terms to zero.
\begin{subequations}
\begin{align}
    P_{i,p,\text{BS}} &\triangleq {{\mathbb{E}\left\{{\left| {{{\bigl[\widehat{\mathbf{\mathbf{\Lambda}}}^{\text{BS}}_{\mathbf{I}}\bigr]_{l^\text{BS}_{s,p},k^\text{BS}_{s,p}}}}} \right|}^2\right\}}} \\
    &= \frac{1}{MN}
    \sum_{\substack{q=1\\q\neq p}}^{K}{\left|\alpha^{\text{BS}}_{s,q,0}\right|^2}{\left|\bigl[\widetilde{\mathbf{\mathbf{\Lambda}}}_{\text{BS}}\bigr]_{(l^\text{BS}_{s,p}-l^\text{BS}_{s,q}),(k^\text{BS}_{s,p}-k^\text{BS}_{s,q})}\right|^2},
\end{align}
\end{subequations}
To quantify the impact of sidelobes produced by unintended targets located at arbitrary positions in the RD map on sensing performance, we examine the integrated sidelobe level (ISL) of the received signal in the RD map after the receiver filter. For the BS, the power of the interference based on ISL at the intended target-$p$ is expressed as
\begin{align}
   P_{i,p,\text{BS}} &=\frac{1}{MN}\sum^{K}_{\substack{q = 1 \\ q\neq p}}
   \left|\alpha^{\text{BS}}_{s,q,0}\right|^2\text{ISL}_{\text{BS}},
\end{align}
where the target-specified information is contained in $\alpha^{\text{BS}}_{s,p,0}$ for all targets, and is formulated as
\begin{subequations}
\begin{align}
\text{ISL}_{\text{BS}}
&= \frac{1}{MN}\Bigg(\sum_{l,k}\bigl|\widetilde{\boldsymbol{\Lambda}}_{\mathrm{BS}}[l,k]\bigr|^{2}
   - \bigl|\widetilde{\boldsymbol{\Lambda}}_{\mathrm{BS}}[0,0]\bigr|^{2}\Bigg) \\
&= \frac{1}{MN}\Bigg(\sum_{m,n} \gamma_{m,n}^{2}
   - \frac{1}{MN}\Big|\sum_{m,n} \gamma_{m,n}\Big|^{2}\Bigg), \label{equ:EP}
\end{align}
\end{subequations} where $\gamma_{m,n} \triangleq \frac{|x_{m,n}|^{2}}{|x_{m,n}|^{2} + \sigma^{2}_{\mathbf{z}_{\mathrm{BS},s}}}$, and the second equality holds because of the energy preservation property of the Fourier transform.

\subsection{Eve Sensing Receiver Processing}

Consistent with the passive sensing literature and according to \textbf{A.5}, the direct path from the BS to the receiver is extracted as a reference, and classical radar processing is performed on the surveillance channel \cite{lingadevaru_analysis_2021, ai_passive_2021, samczynski_5g_2022, abratkiewicz_ssb-based_2023}. Accordingly, Eve filters the TF signal from the surveillance channel, and we assume they use a matched filter as the receive filter\footnote{The MF maximizes the output signal-to-noise ratio (SNR) and requires no knowledge of the CP or frame structure, remaining effective even when the reference-signal SNR is degraded \cite{han_sensing_2025}.}. 

By contrast, alternative mismatched filters generally demand detailed signal-format information, direct-link noise statistics, or direct-link LoS characteristics that are unavailable to Eve \cite{han_sensing_2025}. For both optimal output SNR and implementation simplicity, we therefore assume that Eve adopts the MF, which in the TF domain is given by
\begin{align}
    \left[\widehat{\mathbf{H}}_\text{E,s}\right]_{m,n} &= \left[\mathbf{G}_\text{E}\right]_{m,n} \cdot \left[\mathbf{R}_\text{E,s}\right]_{m,n},
\end{align}
where $\left[\mathbf{G}_\text{E}\right]_{m,n} = x^*_{m,n}$. Observe that $\mathbf{G}_\text{E}$ is the original transmitted signal, as known at the BS, in order to facilitate the analysis, which models Eve’s best-case processing (and therefore the BS’s worst-case). In realistic settings, imperfections in Eve’s reference signal introduce performance loss, consequently degrading their detection performance \cite{han_sensing_2025}.

Similar to the BS, $\widehat{\mathbf{H}}_\text{E,s}$ needs to be transformed into the DD domain to have an RD map for target detection, which can be written as
\begin{align}
    \widehat{\mathbf{\mathbf{\Lambda}}}_\text{E} = \mathbf{F}^H_{M}\widehat{\mathbf{H}}_\text{E,s}\mathbf{F}_{N}.
\end{align}
Further,
\begin{align}
    \bigl[\widehat{\mathbf{\mathbf{\Lambda}}}_{\text{E}}\bigr]_{l,k}
    &= \frac{1}{\sqrt{MN}}
    \sum_{m,n}
    \bigl[\widehat{\mathbf{H}}_{\text{E},s}\bigr]_{m,n}
    e^{\,j2\pi \left(\frac{m\,l}{M} - \frac{n\,k}{N}\right)},
\end{align}
where $l = 0, 1, \cdots, M - 1$ and $k = 0, 1, \cdots, N - 1$. Similarly, we define
\begin{align}
    \bigl[\widehat{\mathbf{\mathbf{\Lambda}}}^{\text{E}}_{\mathbf{S}}\bigr]_{l^\text{E}_{s,p},k^\text{E}_{s,p}}\triangleq&\frac{1}{MN}
    \sum_{m,n}
    \left|x_{m,n}\right|^2 {\alpha^\text{E}_{s,p}},\\
    \bigl[\widehat{\mathbf{\mathbf{\Lambda}}}^{\text{E}}_{\mathbf{I}}\bigr]_{l^\text{E}_{s,p},k^\text{E}_{s,p}}\triangleq&\frac{1}{\sqrt{MN}}
    \sum_{\substack{q=1\\q\neq p}}^{K}{\alpha^{\text{E}}_{s,q}}\bigl[\widetilde{\mathbf{\mathbf{\Lambda}}}_{\text{E}}\bigr]_{(l^\text{E}_{s,p}-l^\text{E}_{s,q}),(k^\text{E}_{s,p}-k^\text{E}_{s,q})},
\end{align}
where $\widetilde{\mathbf{\mathbf{\Lambda}}}_{\text{E}}$ is the AF as Eve applies the MF. The noise component follows
\begin{align}
    \bigl[\widehat{\mathbf{\mathbf{\Lambda}}}^{\text{E}}_{\mathbf{Z}}\bigr]_{l,k}
    &= \frac{1}{\sqrt{MN}}
    \sum_{m,n}
    x^*_{m,n}\bigl[\mathbf{Z}_\text{E,s,TF}\bigr]_{m,n}
    e^{\,j2\pi \left(\frac{m\,l}{M} - \frac{n\,k}{N}\right)},
\end{align}
where $l = 0, 1, \cdots, M - 1$ and $k = 0, 1, \cdots, N - 1$. Similar to the BS, Eve can utilize the mainlobe component to detect and estimate the parameters of target-$p$.

As in the component decomposition in the previous subsection, from (\ref{equ:SINR}) we present the corresponding expressions for the eavesdropper. For Eve, the mainlobe of the desired target-$p$ follows
\begin{align}    
   &P_{s,p,\text{E}} \triangleq {{{\mathbb{E}\left\{\left| {\bigl[\widehat{\mathbf{\mathbf{\Lambda}}}^{\text{E}}_{\mathbf{S}}\bigr]_{l^\text{E}_{s,p},k^\text{E}_{s,p}}} \right|^2\right\}}}} = \frac{\left|\alpha^\text{E}_{s,p,0}\right|^2}{\left(MN\right)^2}
   \left|\sum_{m,n}
   {\lvert x_{m,n}\rvert^2}\right|^2,
\end{align}
and \(\mu_{p,\text{E}}\triangleq\frac{\alpha^{\text{E}}_{s,p,0}}{MN}\sum_{m,n}\lvert x_{m,n}\rvert^2\). The amplified noise power for Eve follows
\begin{align}
    &P_{n,\text{E}} = {\mathbb{E} \left\{ \left| {{\bigl[\widehat{\mathbf{\mathbf{\Lambda}}}^{\text{E}}_{\mathbf{Z}}\bigr]_{l^\text{E}_{s,p},k^\text{E}_{s,p}}}} \right|^2 \right\}} = \frac{\sigma^2_{\mathbf{z}_\text{E,s}}}{MN}\sum_{m,n}
    \lvert x_{m,n}\rvert^2.
\end{align}
The interference produced by other targets for Eve follows
\begin{subequations}
\begin{align}
    P_{i,p,\text{E}} &\triangleq {{\mathbb{E}\left\{{\left| {{{\bigl[\widehat{\mathbf{\mathbf{\Lambda}}}^{\text{E}}_{\mathbf{I}}\bigr]_{l^\text{E}_{s,p},k^\text{E}_{s,p}}}}} \right|}^2\right\}}}\\
    &= \frac{1}{MN}
    \sum_{\substack{q=1 \\ q\neq p}}^{K}{\left|\alpha^{\text{E}}_{s,q,0}\right|^2}{\left|\bigl[\widetilde{\mathbf{\mathbf{\Lambda}}}_{\text{E}}\bigr]_{(l^\text{E}_{s,p}-l^\text{E}_{s,q}),(k^\text{E}_{s,p}-k^\text{E}_{s,q})}\right|^2}.
\end{align}
\end{subequations}
For Eve, the interference power due to the ISL at the intended target-$p$ is given by
\begin{align}
   P_{i,p,\text{E}} &=\frac{1}{MN}\sum^{K}_{\substack{q = 1 \\ q\neq p}}
   \left|\alpha^\text{E}_{s,q,0}\right|^2\text{ISL}_{\text{E}},
\end{align}
where the target-specific information is embedded in $\alpha^\text{E}_{s,p,0}$ for all targets, and the ISL is expressed as
\begin{subequations}
\begin{align}
    \text{ISL}_{\text{E}} &= \frac{1}{MN}\left[\sum_{\substack{l, k}}\left|\left[\widetilde{\mathbf{\mathbf{\Lambda}}}_{\text{E}}\right]_{l,k}\right|^2 - \left|\left[\widetilde{\mathbf{\mathbf{\Lambda}}}_{\text{E}}\right]_{0,0}\right|^2\right] \\
    &= \frac{1}{MN}\left[\sum_{\substack{m,n}}\left|\lvert x_{m,n}\rvert^2\right|^2 - \frac{1}{MN}\left|\sum_{\substack{m,n}}\left| x_{m,n}\right|^2\right|^2\right],
\end{align}
\end{subequations} where the second equality follows from the energy-preserving property of the Fourier transform, similar to (\ref{equ:EP}).

\subsection{KLD Formulation for Sensing-Security Metric}
To characterize detection performance, we adopt the KLD as the metric. Our focus is on secure sensing in ISAC systems, where we aim to maximize the detection KLD at the BS to ensure reliable operation while simultaneously minimizing the detection KLD at Eve to suppress unauthorized sensing.

According to the system model, there are $K$ targets in the scenario. Therefore, the hypothesis test for target detection includes the noise component and sidelobe interference generated by other targets. Because the sidelobes originating from other targets appear in different RD bins yet play statistically identical roles, the central limit theorem justifies modeling their aggregate contribution as a set of i.i.d. zero-mean Gaussian interferences \cite{bedeer_ambiguity_2025}. Consequently, the effective noise variance comprises both ISL and the receiver’s noise. Under this effective noise model, the binary hypothesis test for target detection in a given DD bin $(l,k)$ is formulated as follows.

Taking the detection hypothesis of target-$p$ as an example, we henceforth omit the target index $p$ without loss of generality. For the BS, 
\begin{align}
\bigl[\widehat{\mathbf{\mathbf{\Lambda}}}_{\text{BS}}\bigr]_{l,k}
  &= 
  \begin{cases}
      \bigl[\widehat{\mathbf{\mathbf{\Lambda}}}^{\text{BS}}_{\mathbf{I}}\bigr]_{l,k}+\bigl[\widehat{\mathbf{\mathbf{\Lambda}}}^{\text{BS}}_{\mathbf{Z}}\bigr]_{l,k}, 
      &\mathcal{H}_{0,\text{BS}},\\[4pt]
      \bigl[\widehat{\mathbf{\mathbf{\Lambda}}}^{\text{BS}}_{\mathbf{S}}\bigr]_{l,k}
      + \bigl[\widehat{\mathbf{\mathbf{\Lambda}}}^{\text{BS}}_{\mathbf{I}}\bigr]_{l,k}
      +\bigl[\widehat{\mathbf{\mathbf{\Lambda}}}^{\text{BS}}_{\mathbf{Z}}\bigr]_{l,k}, 
      &\mathcal{H}_{1,\text{BS}},
  \end{cases}
\end{align}
where $l=0, \cdots, M-1$ and $k=0, \cdots, N-1$. And for Eve,
\begin{align}
\bigl[\widehat{\mathbf{\mathbf{\Lambda}}}_{\text{E}}\bigr]_{l,k}
  &= 
  \begin{cases}
      \bigl[\widehat{\mathbf{\mathbf{\Lambda}}}^{\text{E}}_{\mathbf{I}}\bigr]_{l,k}
      +\bigl[\widehat{\mathbf{\mathbf{\Lambda}}}^{\text{E}}_{\mathbf{Z}}\bigr]_{l,k}, 
      &\mathcal{H}_{0,\text{E}},\\[4pt]
      \bigl[\widehat{\mathbf{\mathbf{\Lambda}}}^{\text{E}}_{\mathbf{S}}\bigr]_{l,k}
      +
      \bigl[\widehat{\mathbf{\mathbf{\Lambda}}}^{\text{E}}_{\mathbf{I}}\bigr]_{l,k}
      + \bigl[\widehat{\mathbf{\mathbf{\Lambda}}}^{\text{E}}_{\mathbf{Z}}\bigr]_{l,k}, 
      &\mathcal{H}_{1,\text{E}},
  \end{cases}
\end{align}
where $l=0, \cdots, M-1$ and $k=0, \cdots, N-1$. 
Therefore, using the norm-squared filtered signal with the generalized likelihood ratio test (GLRT) detector\footnote{In the simulations, the cell-averaging constant false alarm rate (CA-CFAR) detector, which can be regarded as a general extension of the GLRT, is used when the effective noise power is unknown to the detector.}, the received signal in the RD map is defined as $T_\text{BS} \triangleq \left|\bigl[\widehat{\mathbf{\mathbf{\Lambda}}}_{\text{BS}}\bigr]_{l,k}\right|^2$ and $T_\text{E} \triangleq \left|\bigl[\widehat{\mathbf{\mathbf{\Lambda}}}_{\text{E}}\bigr]_{l,k}\right|^2$.

\begin{figure}
    \centering
    \includegraphics[width=\linewidth]{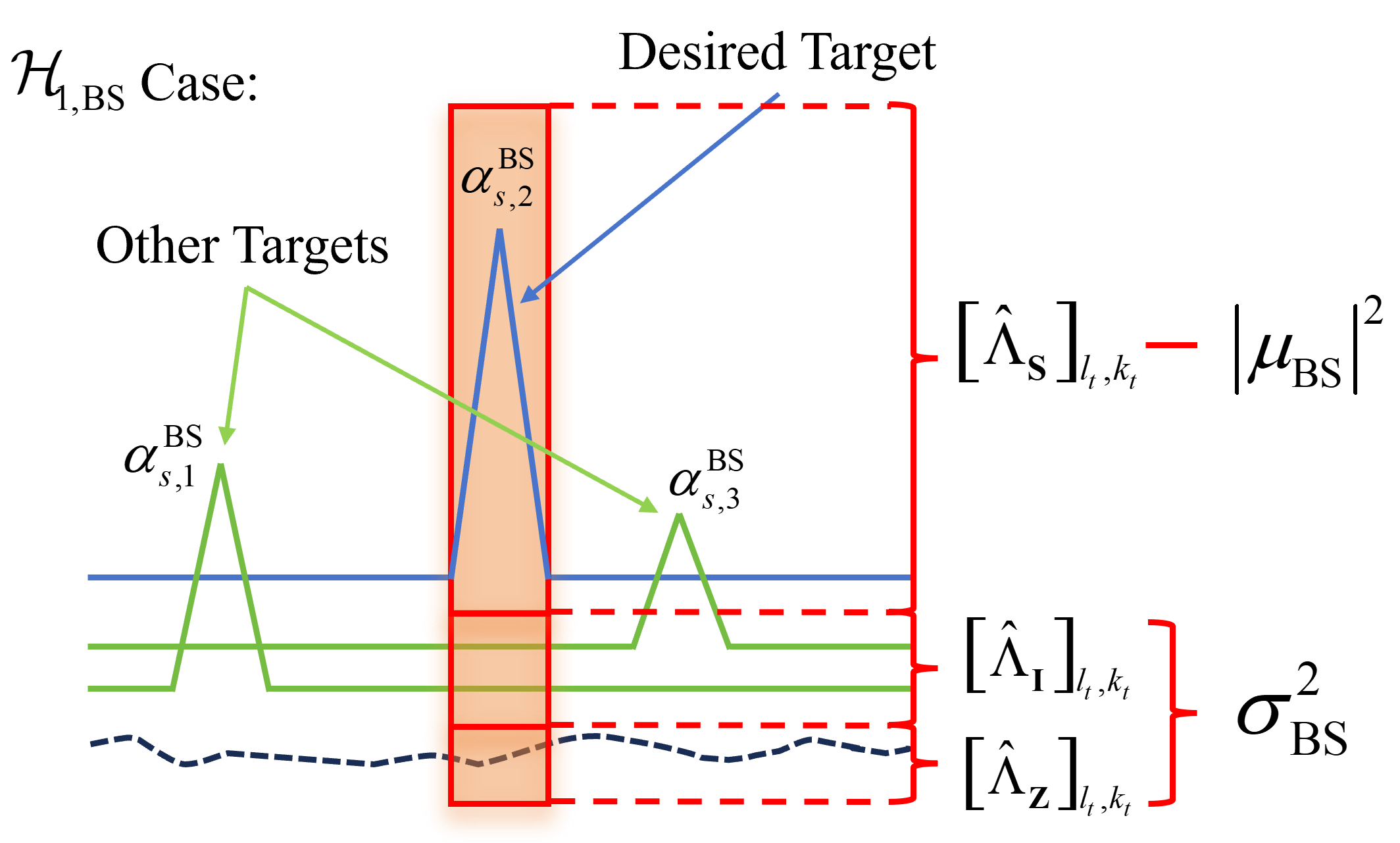}
    \caption{KLD components for $\mathcal{H}_{1,\mathrm{BS}}$ at the BS: filtered RD-domain signal of the desired target (blue), other targets (green), and noise level (black dashed).}
    \label{fig:KLD_illu}
\end{figure}

Given the i.i.d. Gaussian assumption of the ISL from different targets, the probability distribution of the received signal at the BS's and Eve's sides can be expressed as follows using exponential distributions. In the $\mathcal{H}_0$ case:
\begin{align}
     T_\text{BS} \sim \text{Exp}\left(\sigma^2_{\text{BS}}\right), T_\text{E} \sim \text{Exp}\left(\sigma^2_{\text{E}}\right), 
\end{align}
where $\sigma^2_\text{BS}\triangleq P_{i,p,\text{BS}} + P_{n,\text{BS}}$, and $\sigma^2_\text{E}\triangleq P_{i,p,\text{E}} + P_{n,\text{E}}$ are the effective noise variances for BS and Eve, respectively. Their PDFs are
\begin{align}
f_0\left(T_\text{BS}\right) &= \frac{1}{\sigma^2_{\text{BS}}}\operatorname{exp}\left(-\frac{T_\text{BS}}{\sigma^2_{\text{BS}}}\right),\\
f_0\left(T_\text{E}\right) &= \frac{1}{\sigma^2_{\text{E}}}\operatorname{exp}\left(-\frac{T_\text{E}}{\sigma^2_{\text{E}}}\right), 
\end{align}
where $T_\text{BS} \geq 0$ and $T_\text{E} \geq 0$. In the $\mathcal{H}_1$ case:
\begin{align}
    \frac{2}{\sigma_{\text{BS}}^{2}}\,T_{\text{BS}} &\sim 
    \chi^{\prime 2}_{2}\!\left(\frac{2|\mu_{\text{BS}}|^{2}}{\sigma_{\text{BS}}^{2}}\right),
    \frac{2}{\sigma_{\text{E}}^{2}}\,T_{\text{E}} \sim 
    \chi^{\prime 2}_{2}\!\left(\frac{2|\mu_{\text{E}}|^{2}}{\sigma_{\text{E}}^{2}}\right),
\end{align}
where $\chi^{\prime 2}_{2}(\cdot)$ denotes the non-central chi-square distribution
with two degrees of freedom, $\mu_{\text{BS}} = \mu_{p,\text{BS}}$ and $\mu_{\text{E}} = \mu_{p,\text{E}}$. Fig.~\ref{fig:KLD_illu} illustrates the KLD components at the BS under $\mathcal{H}_{1,\mathrm{BS}}$: the effective noise ($\sigma^2_{\text{BS}}$) comprises noise and interference, and the mean ($\mu_{\text{BS}}$) corresponds to the desired target’s mainlobe. 
An analogous description applies to Eve under $\mathcal{H}_{1,\mathrm{E}}$. Accordingly, their probability distributions are
\begin{align}
    f_1\left(T_\text{BS}\right) &= \frac{1}{\sigma^2_{\text{BS}}}\operatorname{exp}\left(-\frac{T_\text{BS} + \left|\mu_{\text{BS}}\right|^2}{\sigma^2_{\text{BS}}}\right)I_0\left(\frac{2\left|\mu_{\text{BS}}\right|\sqrt{T_\text{BS}}}{\sigma^2_{\text{BS}}}\right),\\
    f_1\left(T_\text{E}\right) &= \frac{1}{\sigma^2_{\text{E}}}\operatorname{exp}\left(-\frac{T_\text{E} + \left|\mu_{\text{E}}\right|^2}{\sigma^2_{\text{E}}}\right)I_0\left(\frac{2\left|\mu_{\text{E}}\right|\sqrt{T_\text{E}}}{\sigma^2_{\text{E}}}\right),
\end{align}
where $T_\text{BS} \geq 0$, $T_\text{E} \geq 0$, and both follow non-central chi-square distributions. Therefore, the KLD for the BS can be written as
\begin{subequations}
\label{eq:kl_bs}
\begin{align}
D_{\text{BS}}
&=\int_{0}^{\infty}
   f_1(T)\,
   \ln\frac{f_1(T)}{f_0(T)}
   \,\text{d}T
\label{eq:kl_bs_def}\\
&=\int_{0}^{\infty}
   \frac{1}{\sigma^2_{\text{BS}}}
   \exp\!\Bigl(
     -\frac{T + |\mu_{\text{BS}}|^2}
           {\sigma^2_{\text{BS}}}
   \Bigr)
   I_0\!\Bigl(
     \frac{2|\mu_{\text{BS}}|\sqrt{T}}
          {\sigma^2_{\text{BS}}}
   \Bigr)\notag\\[-3pt]
&\qquad\times
   \ln\Biggl[\exp\!\Bigl(-\tfrac{|\mu_{\text{BS}}|^2}
                      {\sigma^2_{\text{BS}}}\Bigr)
     I_0\!\Bigl(
       \tfrac{2|\mu_{\text{BS}}|\sqrt{T}}
             {\sigma^2_{\text{BS}}}
     \Bigr)
   \Biggr]
   \,\text{d}T
\label{eq:kl_bs_sub}\\
&=\mathbb{E}_{T\sim f_1}\!
   \Bigl[
     \ln I_0\!\bigl(
       \tfrac{2|\mu_{\text{BS}}|\sqrt{T}}
             {\sigma^2_{\text{BS}}}
     \bigr)
   \Bigr]
  -\frac{|\mu_{\text{BS}}|^2}
        {\sigma^2_{\text{BS}}}.
\label{eq:kl_bs_result}
\end{align}
\end{subequations}
The KLD for Eve has a similar expression as
\begin{align}
    D_{\text{E}}(\mathcal{H}_1\parallel \mathcal{H}_0)
    &=\mathbb{E}_{T\sim f_1}\!
   \Bigl[
     \ln I_0\!\bigl(
       \tfrac{2|\mu_{\text{E}}|\sqrt{T}}
             {\sigma^2_{\text{E}}}
     \bigr)
   \Bigr]
  -\frac{|\mu_{\text{E}}|^2}
         {\sigma^2_{\text{E}}}.
\end{align}
Note that, $D_{\text{BS}}(\mathcal{H}_1\parallel \mathcal{H}_0)$ and $D_{\text{E}}(\mathcal{H}_1\parallel \mathcal{H}_0)$ are functions of $\left[w_1, w_2, \cdots, w_{MN}\right]^T$, given a specific data sequence. Consequently, secure sensing can be achieved by maximizing the KLD at the BS while minimizing it at Eve. Additionally, the derived KLD can be approximated for better tractability as
\begin{align}
\widetilde{D}_{\text{BS}}
&\triangleq - \frac{|\mu_{\text{BS}}|^{2}}{\sigma_{\text{BS}}^{2}} + \ln\Biggl\{
  I_{0}\Biggl[
    \frac{2|\mu_{\text{BS}}|}{\sigma_{\text{BS}}^{2}}\,
    \sqrt{\sigma_{\text{BS}}^{2}}\,
    \frac{\sqrt{\pi}}{2}\,
    \exp\!\Bigl(-\frac{|\mu_{\text{BS}}|^{2}}{2\sigma_{\text{BS}}^{2}}\Bigr) \notag\\
    &\times\Bigl(
      \Bigl(1+\frac{|\mu_{\text{BS}}|^{2}}{\sigma_{\text{BS}}^{2}}\Bigr)
      I_{0}\!\Bigl(\frac{|\mu_{\text{BS}}|^{2}}{2\sigma_{\text{BS}}^{2}}\Bigr)
      + \frac{|\mu_{\text{BS}}|^{2}}{\sigma_{\text{BS}}^{2}}
      I_{1}\!\Bigl(\frac{|\mu_{\text{BS}}|^{2}}{2\sigma_{\text{BS}}^{2}}\Bigr)
    \Bigr)
  \Biggr]
\Biggr\},
\end{align}
which is detailed in Appendix \ref{app:KLD_approx}. For the eavesdropper case, the approximated KLD is obtained by rewriting the expressions in terms of Eve’s parameters. Finally, we define the KLD gap as $\frac{\widetilde{D}_{\text{BS}}}{\widetilde{D}_{\text{E}}}$ for the sensing-security metric.

\section{Sensing-Secured ISAC Optimization}
\label{sec:opt_prob}
In this section, we formulate the optimization problem for the sensing-secure ISAC signaling design, followed by the optimization method that solves the proposed problem using the simulated annealing algorithm.

\subsection{Optimization Problem}
As discussed previously, securing the sensing functionality requires maximizing BS's KLD and minimizing Eve’s KLD, thereby making the hypothesis test more distinguishable at the BS and less distinguishable at Eve. Consequently, the optimization problem for ISAC sensing-security can be formulated as follows
\begin{subequations}\label{eq:optimization}
\begin{align}
\underset{\mathbf{W}_\text{TF}}{\text{max}}\quad
& \beta \left[\frac{\widetilde{D}_{\text{BS}}\left(\mathbf{W}_\text{TF}\right)}{\widetilde{D}_{\text{E}}\left(\mathbf{W}_\text{TF}\right)}\right]
  - (1 - \beta)\,{{{\left\| {{{\mathbf{H}}_{{\text{c}},{\text{eff}}}}{{\mathbf{W}}_{{\text{TF}}}}{\mathbf{Us}} - {\mathbf{s}}} \right\|}^2_2}}
  \label{eq:opt_obj}\\
\text{s.t.\ } \quad &\operatorname{Tr}\left(\mathbf{W}^H_\text{TF}\mathbf{W}_\text{TF}\right) \le P_{\max},
  \label{eq:opt_power}
\end{align}
\end{subequations}
where $\beta$ is the trade-off weighting factor, and $\mathbf{W}_\text{TF}$ is the TF perturbation matrix, providing joint amplitude and phase control as mentioned previously. Note that this optimization is formulated for a specific data sequence, i.e., a data-sequence-level optimization, and assumes the communication channel matrix is known at the BS. 

\begin{remark}
We note that the above optimization allows for both Eve-aware and Eve-agnostic designs. Indeed, $\widetilde{D}_{\text{E}}$ is defined in terms of Eve’s channel coefficients and noise power. If these quantities are known to the BS, this allows an Eve-aware approach. In practical implementations, however, the BS does not always have Eve’s exact parameters: these parameters can be assigned arbitrary values, and the proposed method therefore remains applicable in Eve-agnostic scenarios. Evidently, in this case, the achievable security performance may result in a different sensing-security--communication trade-off than the ideal Eve-aware case, while any additional information about Eve, if available, can be incorporated into these parameters.
\end{remark}

\begin{remark}
Moreover, $\widetilde{D}_{\text{E}}$ is evaluated under the ideal assumption of a perfect reference signal at Eve, and thus represents an optimistic case on Eve’s sensing performance and a conservative (worst-case) scenario from the BS perspective.
\end{remark}

\subsection{Optimization Solution}

The problem \eqref{eq:optimization} is non-convex and exhibits a complex structure. Although the approximated KLDs, $\widetilde{D}_{\text{BS}}$ and $\widetilde{D}_{\text{E}}$, are convex with respect to (w.r.t.) $\frac{|\mu_{\text{BS}}|^{2}}{\sigma^2_{\text{BS}}}$ and $\frac{|\mu_{\text{E}}|^{2}}{\sigma^2_{\text{E}}}$, respectively, the intricate functional dependencies of $\mu_{\text{BS}}$, $\mu_{\text{E}}$, $\sigma^2_{\text{BS}}$, and $\sigma^2_{\text{E}}$ on $\left[w_1, \cdots, w_{MN}\right]^T$ further complicate the structure of problem \eqref{eq:optimization}. Therefore, we adopt the simulated annealing algorithm to solve the problem (\ref{eq:optimization}).

The simulated annealing algorithm is a well-established optimization method for tackling nonconvex problems and obtaining high-quality approximate solutions \cite{ingber_1993_simulated, delahaye_2019_simulated}. Starting from an initial solution, simulated annealing iteratively generates candidate solutions and, under a gradually decreasing temperature parameter, accepts improving moves while occasionally accepting deteriorating ones, thereby escaping local optima and enhancing global exploration. As the temperature is lowered, the search gradually shifts from global exploration to local exploitation, and the algorithm typically converges to a near-optimal solution \cite{delahaye_2019_simulated}. Convex optimization may be a promising area for future research in solving this problem.

\newcommand{\xbs}{P_{n,\text{BS}}}
\newcommand{\ybs}{\text{ISL}_{\text{BS}}}
\newcommand{\sbs}{P_{s,\text{BS}}}
\newcommand{\xe}{P_{n,\text{E}}}
\newcommand{\ye}{\text{ISL}_{\text{E}}}
\newcommand{\se}{P_{s,\text{E}}}

\section{Numerical Results}
\label{sec:Num_res}

In this section, we present numerical results to validate the proposed sensing-security scheme for ISAC. Firstly, under the Eve-aware scenario, the detection probability for BS and Eve is evaluated, and illustrations comparing the AFs with and without the sensing-security design are provided. In addition, a sensing case is investigated to further demonstrate the security effect by showing the different RD maps of BS and Eve. Subsequently, the trade-off between sensing-security and communication performance is assessed under different clutter environments, illustrating how the optimization allocates resources between the two functionalities in Eve-aware scenarios. Lastly, we present the comparison of the proposed method for Eve-aware and Eve-agnostic cases.

OTFS or OFDM with a given data sequence is assumed to be transmitted by the ISAC BS, while the sensing eavesdropper first extracts the reference signal from their direct link, equalizes it to estimate the transmitted signal, and then performs a passive sensing process. Unless stated otherwise, the communication information is modulated with the 16QAM constellation. The number of Monte Carlo realizations for symbol sequences and noise is set to 100. Detection is implemented using CA-CFAR with a constant threshold of 10 dB. Other experimental setups are summarized in Table \ref{tab:SS_simu_params} and \ref{tab:TO_simu_params} for the first two experiments, respectively.

\subsection{Secure-Sensing Performance Evaluation: Eve-Aware Case}
\label{sec:SS_SNR}
\begin{table}[t!]
    \caption{Simulation parameters for sensing-security centric design}
    \label{tab:SS_simu_params}
    \centering
    \setlength{\tabcolsep}{6pt}
    \renewcommand{\arraystretch}{1.15}
    \begin{tabular}{@{}l c@{}}
        \toprule
        \textbf{Parameter} & \textbf{Value} \\
        \midrule
        ISAC balance weight $\beta$ & 1 \\
        Number of subcarriers $M$ & 8 \\
        Number of time-slots $N$ & 8 \\
        Communication channel delay $l_\text{c}$ & 2 \\
        Communication channel Doppler $k_\text{c}$ & 3 \\
        Communication channel coefficient $\alpha_\text{c}$ & 1 \\
        Communication SNR $SNR_\text{c}$ & 25 dB \\
        Communication Rician factor $\kappa_\text{c}$ & 10 dB \\
        Number of targets $K$ & 3 \\
        Unintended-target coeff. range $\alpha^{\text{BS}}_{\text{s},q,0}$, $\alpha^{\text{E}}_{\text{s},q,0}$ & -15–10 dB \\
        Direct link SNR for Eve $SNR_\text{ref}$ & 0 dB \\
        Direct link Rician factor for Eve $\kappa_\text{ref}$ & 0 dB \\
        \bottomrule
    \end{tabular}
\end{table}

In this subsection, we present experimental results for the detection probability and KLD of BS and Eve, with and without the security design, to further demonstrate the sensing-security performance and its relationship with KLD. In addition, to highlight the key manipulations in the sensing-security design, we present the AFs of the transmitted signal. Both OTFS and OFDM waveforms are applied to evaluate the performance. Channel coefficients of unintended targets decrease uniformly across the specified range (as shown in Table \ref{tab:SS_simu_params}).

\begin{figure*}[!t]
  \centering
  \subfloat[Detection probability curves of BS and Eve for different sensing schemes.\label{fig:Pd_curves_a}]{
      \includegraphics[width=0.95\linewidth]{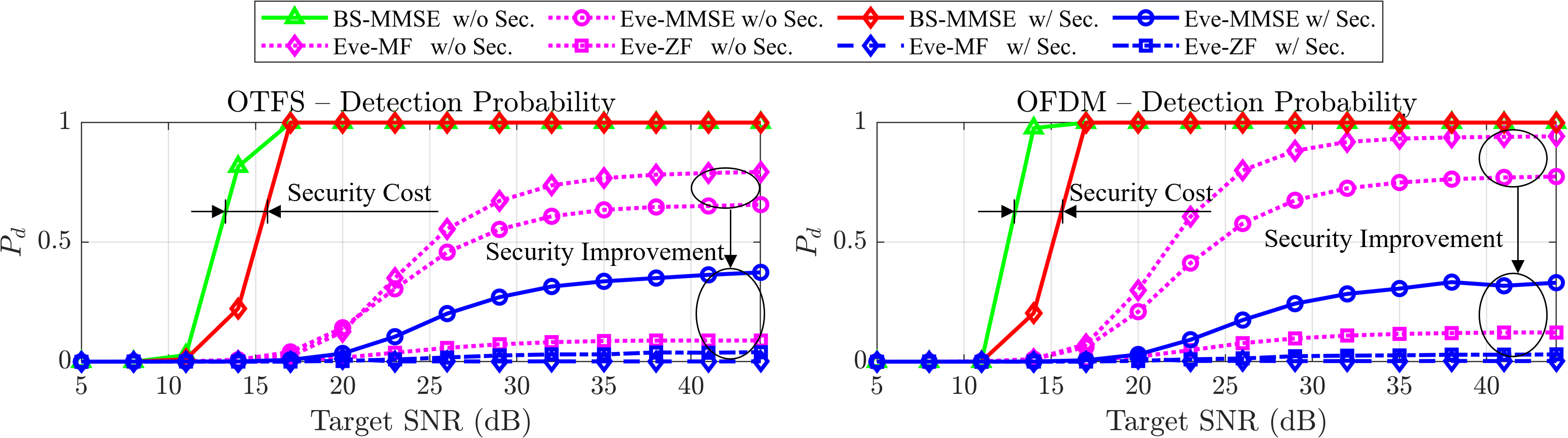}
  }\hfill
  \subfloat[KLD curves of BS and Eve for different sensing schemes.\label{fig:Pd_curves_b}]{
      \includegraphics[width=0.95\linewidth]{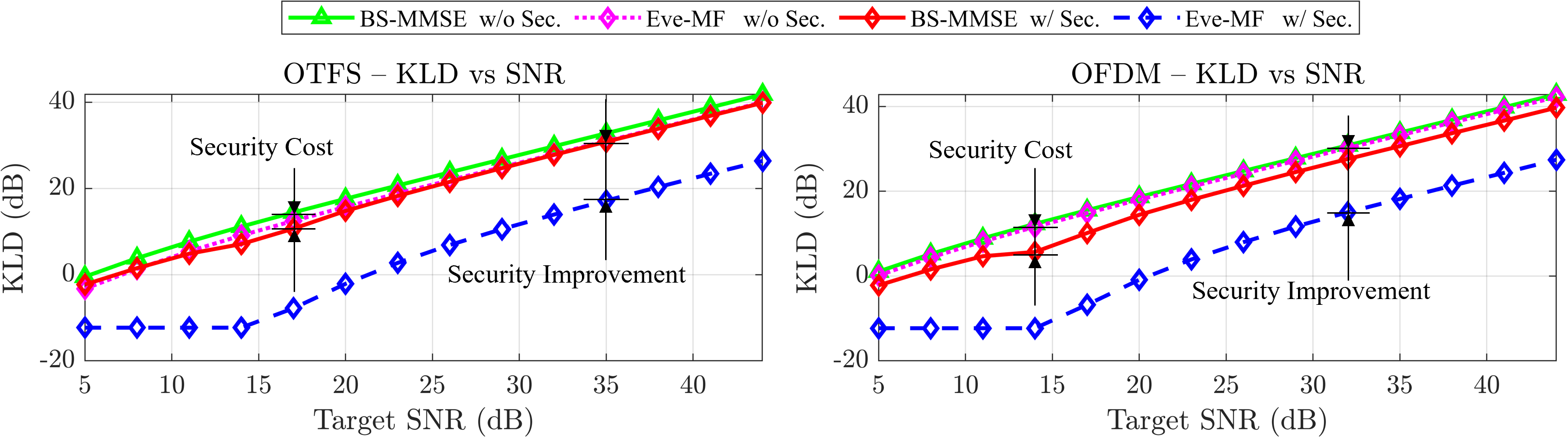}
  }
  \caption{Detection probability and KLD curves of BS and Eve for different sensing schemes, sensing-security-centric design ($\beta=1$).}
  \label{fig:Pd_curves}
\end{figure*}

The detection probability curves are illustrated in Fig.~\ref{fig:Pd_curves_a}. It can be seen that the optimization algorithm effectively obtains the optimized solution of the sensing-security problem~(\ref{eq:optimization}). The optimized results of the detection probability curves for both OTFS and OFDM waveforms demonstrate that the proposed sensing-security design effectively suppresses Eve’s detection probability, albeit with a slight offset in the detection probability of the BS. It is worth noting that, before the security design, the MF is the best filter for Eve, and it is the one most strongly suppressed by the proposed sensing-security processing. In addition, not only the MF, but also other filters, e.g., zero-forcing (ZF) and LMMSE, are suppressed by this method. Meanwhile, the optimized KLD curves in Fig.~\ref{fig:Pd_curves_b} also show that the proposed design effectively suppresses the KLD of Eve, albeit with a slight decrease in that of the BS.

\begin{figure*}[ht]
    \centering
    \includegraphics[width=0.9\linewidth]{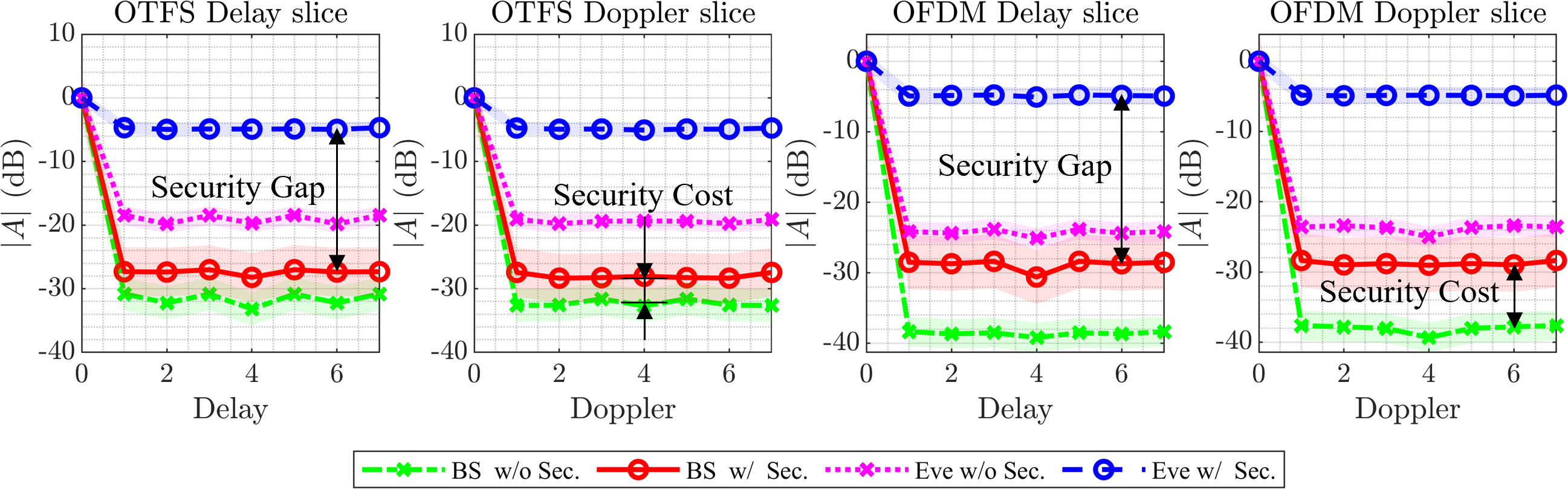}
    \caption{Averaged ambiguity functions of BS and Eve when target SNR is at 10 dB. The shadow areas qualitatively reflect the noise power for the corresponding methods.}
    \label{fig:AF_10dB}
\end{figure*}

Fig.~\ref{fig:AF_10dB} illustrates the AFs of the transmitted signal at a representative target SNR of $10$~dB, showing delay and Doppler slices for both OTFS and OFDM, with and without the proposed sensing-security optimization. The optimal solution consistently raises Eve's sidelobe floor for both waveforms, while the BS sidelobes remain clustered at much lower levels, with only a modest increase compared to the non-secure case. This vertical separation between the BS and Eve curves, highlighted as the ``security gap'' in the figure, can be interpreted as an AF-based security margin: the BS retains a low-sidelobe AF conducive to reliable detection, whereas Eve observes a significantly more distorted AF with elevated sidelobes. Consequently, although the waveform is designed via a KLD-based criterion, its practical effect is to raise Eve’s effective interference level (or equivalently, the effective noise variance in the KLD), thereby degrading Eve's detection probability while incurring only a limited sensing-performance cost at the BS.

\begin{figure*}[ht]
    \centering
    \includegraphics[width=\linewidth]{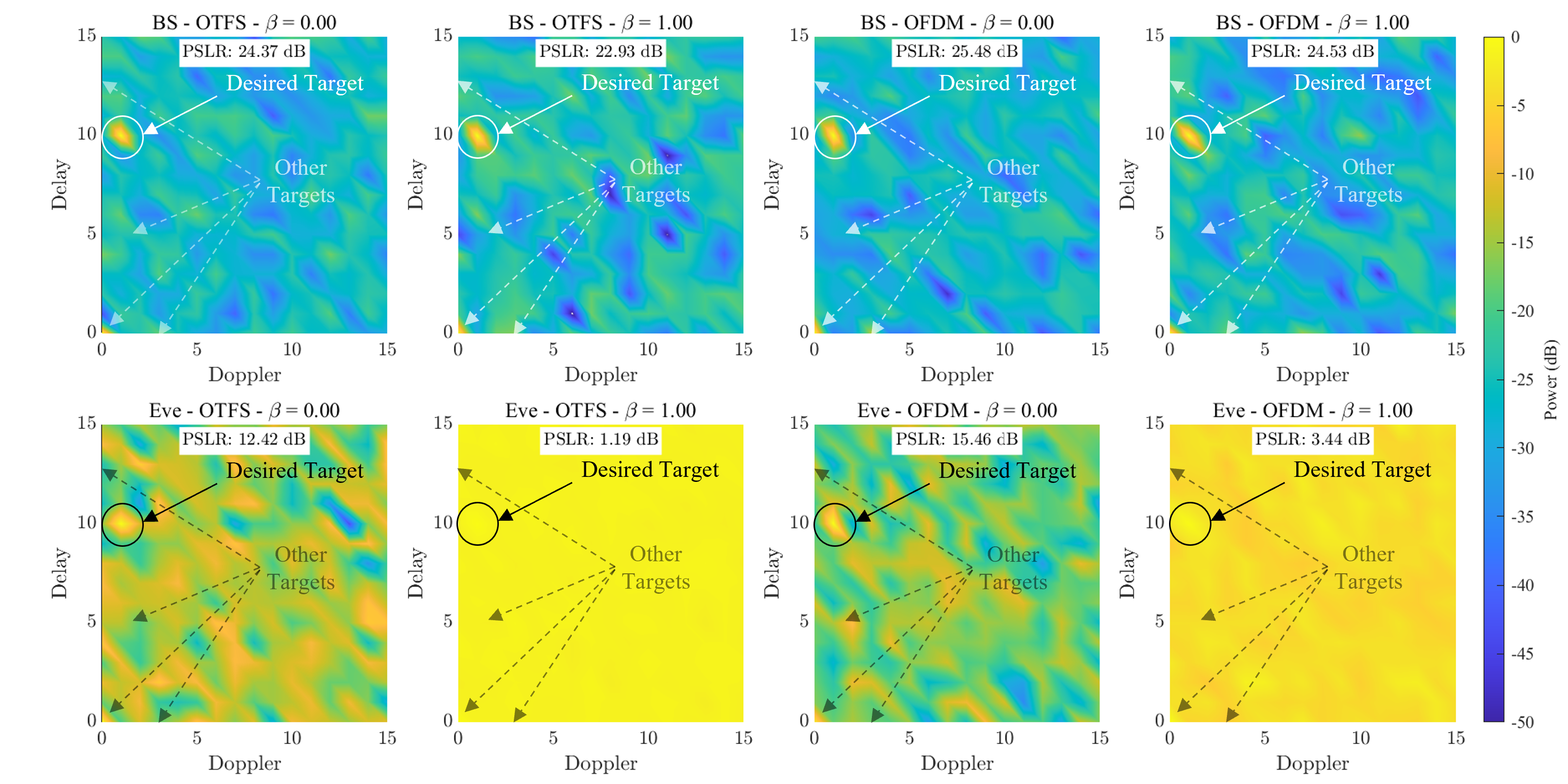}
    \caption{RD maps versus the sensing-security preference for BS (top) and Eve (bottom), using optimized sensing-secured ISAC signals at a target SNR of 20 dB; the numbers of subcarriers and time slots, $M$ and $N$, are 16; the number of targets, $K$, is 5.}
    \label{fig:case_show}
\end{figure*}

Fig.~\ref{fig:case_show} illustrates the RD maps of a specific case for the BS and Eve under different weighting preference factors, $\beta$, using a specific data sequence to transmit. For the communication-centric design ($\beta = 0$), the desired target exhibits high RD contrast at both nodes: the peak-to-sidelobe level ratio (PSLR) for the desired target at the BS is $24.37~\text{dB}$ and $25.48~\text{dB}$ for OTFS and OFDM, respectively, while Eve attains PSLR values of $12.42~\text{dB}$ (OTFS) and $15.46~\text{dB}$ (OFDM), indicating that the target remains relatively well resolved. In contrast, for the sensing-security-centric design ($\beta = 1$), the BS still preserves a pronounced mainlobe with PSLR of $22.93~\text{dB}$ (OTFS) and $24.53~\text{dB}$ (OFDM), whereas Eve's PSLR is drastically reduced to $1.19~\text{dB}$ and $3.44~\text{dB}$, respectively, rendering the desired target barely distinguishable from the surrounding clutter. These quantitative results substantiate that the proposed waveform design impairs Eve's sensing capability while maintaining reliable target detection at the BS.

\begin{figure*}[ht]
    \centering
    \includegraphics[width=\linewidth]{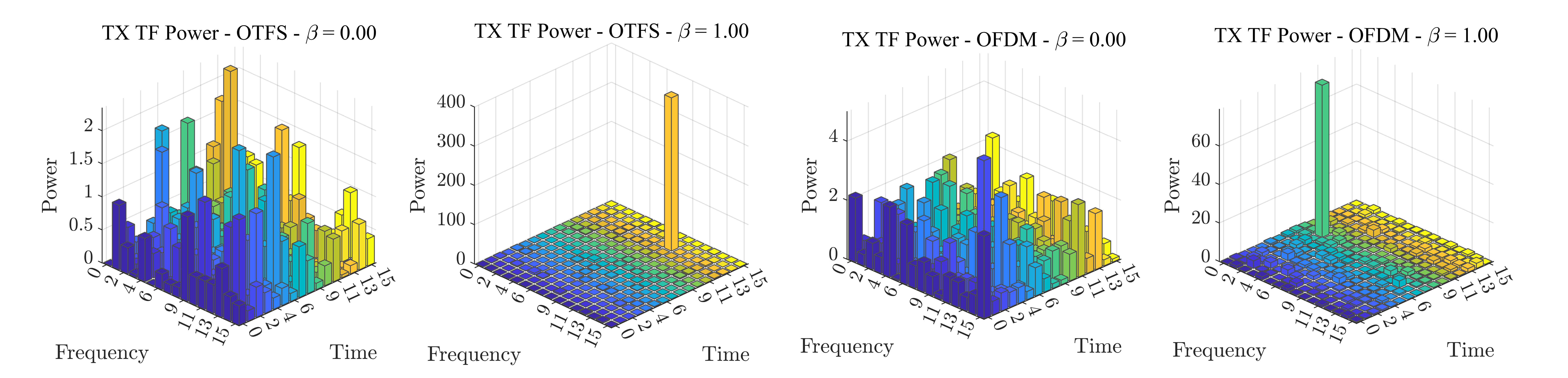}
    \caption{TF-domain power distribution of the transmitted signal versus security preference for OTFS and OFDM at target SNR = 20 dB; the numbers of subcarriers and time slots, $M$ and $N$, are 16; the number of targets, $K$, is 5.}
    \label{fig:TF}
\end{figure*}

Fig.~\ref{fig:TF} illustrates the TF-domain power distributions of the transmitted OTFS and OFDM waveforms under $\beta=0$ and $\beta=1$. For $\beta=0$, the transmit power is broadly spread over the TF grid, which is consistent with a communication-centric design. For $\beta=1$, the optimization yields a highly concentrated TF power pattern, reflecting a clear trade-off between communication performance and sensing-security. Importantly, this concentration at the BS does not impair its sensing performance: under the considered high-SNR regime, the BS LMMSE processor operates close to a ZF receiver, effectively cancelling the contributions of all transmit symbols. Hence, even though many TF samples exhibit low power, they still carry channel information that can be recovered after ZF, preserving the BS’s RD response structure. In contrast, Eve relies on an MF receiver, for which the concentrated transmit power leads to significant sidelobes in the RD map, thereby increasing interference and degrading Eve’s detection performance. In summary, communication favors spread randomness, sensing favors spread determinism \cite{liu_deterministic_2023}, whereas sensing-security favors concentrated determinism in the TF domain as shown in Fig.~\ref{fig:TF}.

\subsection{Sensing-Secure ISAC Trade-off: Eve-Aware Case}

\begin{table}[t!]
    \caption{Simulation parameters for the sensing-secured ISAC trade-off experiment}
    \label{tab:TO_simu_params}
    \centering
    \setlength{\tabcolsep}{6pt}
    \renewcommand{\arraystretch}{1.15}
    \begin{tabular}{@{}l c@{}}
        \toprule
        \textbf{Parameter} & \textbf{Value} \\
        \midrule    
        Number of subcarriers $M$ & 8 \\
        Number of time-slots $N$ & 8 \\
        Communication channel delay $l_\text{c}$ & 2 \\
        Communication channel Doppler $k_\text{c}$ & 3 \\
        Communication channel coefficient $\alpha_\text{c}$ & 1 \\
        Communication SNR $SNR_\text{c}$ & 25 dB \\
        Communication Rician factor $\kappa_\text{c}$ & 10 dB \\
        Number of targets $K$ & 3 \\
        Target SNR for BS and Eve $SNR_\text{s}$ & 25 dB \\
        Direct link SNR for Eve $SNR_\text{ref}$ & 20 dB \\
        Direct link Rician factor for Eve $\kappa_\text{ref}$ & 10 dB \\
        \bottomrule
    \end{tabular}
\end{table}

In this subsection, we present results to illustrate the trade-off between sensing-security and communication performance. Specifically, we investigate the bottom-end metrics for sensing-security and communication—detection probability and average communication rate—under various clutter environments.

\begin{figure}[ht]
  \centering
  \subfloat[Detection probability trade-off transition process under different scatter reflection conditions.
    \label{fig:Pd_TO}]{
      \includegraphics[width=\linewidth]{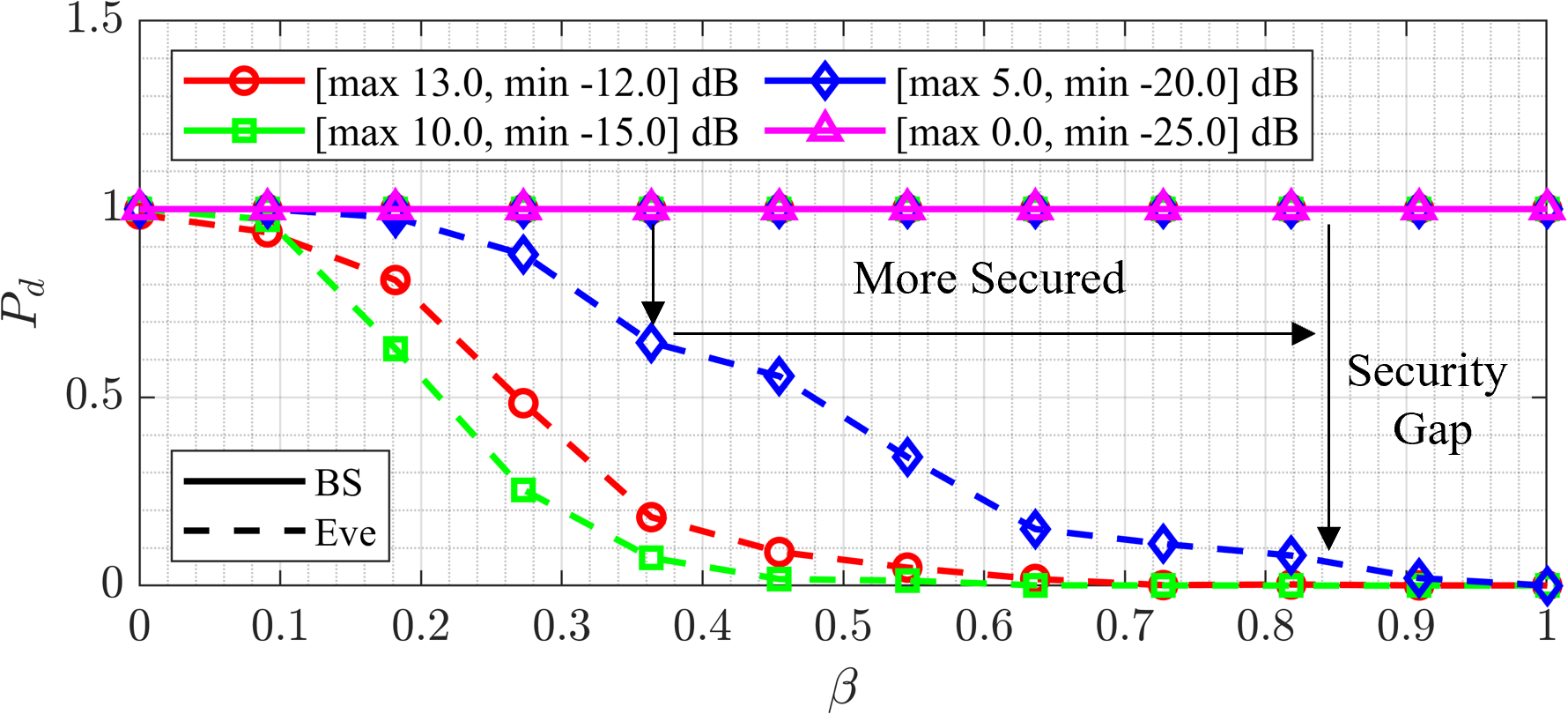}
  }\hfill
  \subfloat[Average communication rate trade-off transition process under different scatter reflection conditions.
    \label{fig:Rate_TO}]{
      \includegraphics[width=\linewidth]{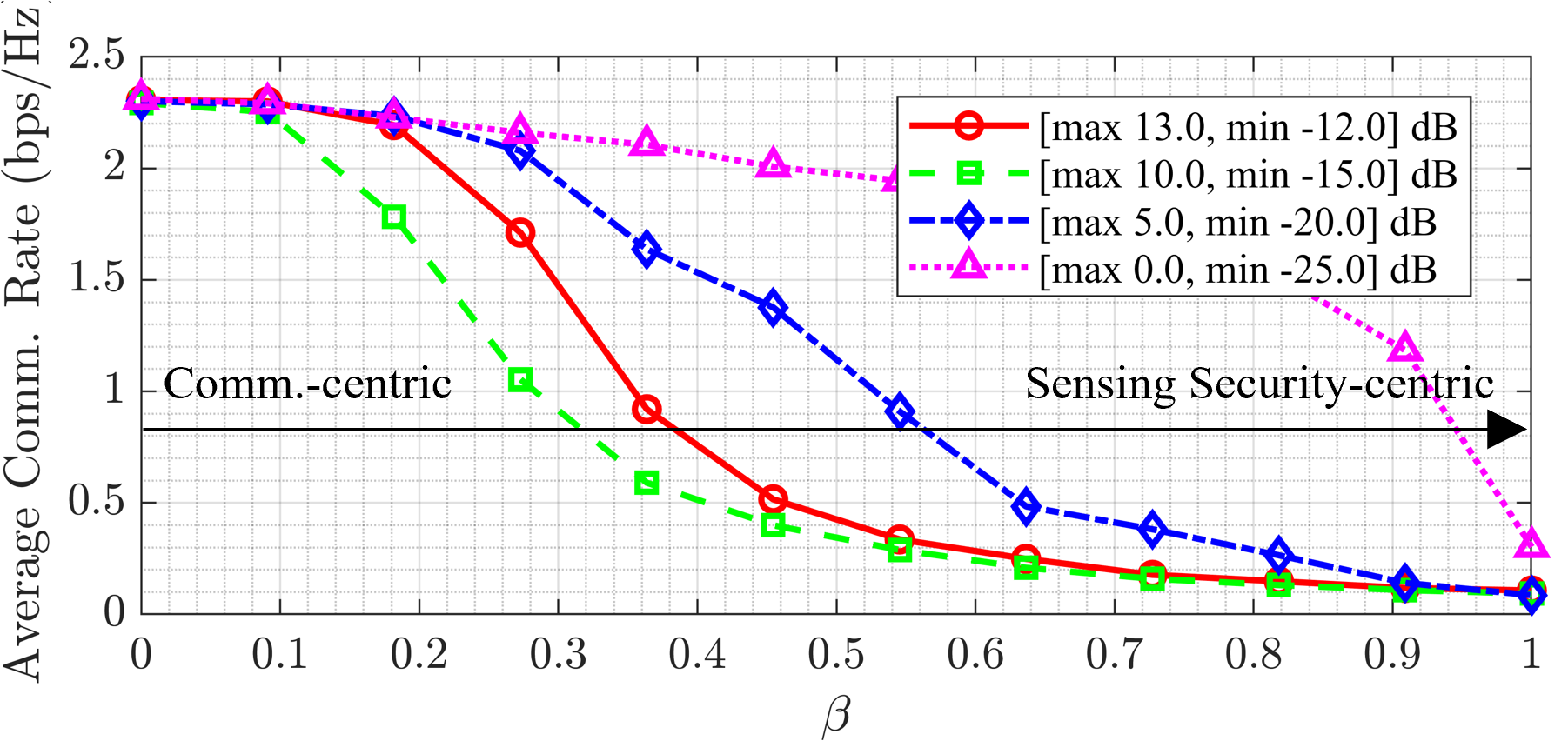}
  }
  \caption{The trade-off transition process for both sensing-security and communication metrics for OTFS.}
  \label{fig:TO_process}
\end{figure}

Fig.~\ref{fig:TO_process} shows the trade-off. The desired and other targets are considered in an environment with channel coefficients spanning a specified range and decreasing in uniform steps from the upper to the lower bound, as labeled in the figure. In Fig.~\ref{fig:Pd_TO}, as priority weight $\beta$ increases from 0 to 1, the gap between the detection probabilities of BS and Eve widens: the detection probability of Eve decreases, while that of BS remains essentially unchanged. However, when the scatter-reflection power is sufficiently low, the security effect reflected in the detection probability vanishes, indicating that guaranteeing sensing-security is challenging in near scatter-free environments. Since scatterers are ubiquitous in practice, this approach remains practical for sensing-security. In addition, under the most severe clutter conditions ($\text{[-12, 13]}$ dB), security performance begins to improve later than in the $\text{[-15, 10]}$ dB case, indicating that, with severe clutter, the system can still achieve a secure design but finds it harder to enhance the BS sensing performance while degrading Eve’s. As a consequence, sensing-security is difficult to realize under both extremely severe clutter and less-cluttered conditions, reflecting an inherent conflict between the BS and Eve: severe clutter degrades Eve but also the BS, whereas less clutter improves the BS and likewise Eve; however, with proper exploitation of clutter, sensing-security can still be achieved. Fig.~\ref{fig:Rate_TO} illustrates the communication performance under different priority weights and clutter conditions, indicating that the security performance comes at the expense of communication degradation. In both less-cluttered and severely cluttered environments, the available degrees of freedom cannot be effectively leveraged to enhance sensing-security—owing to the sensing-performance conflict between BS and Eve—so the optimization naturally prioritizes communication performance. Consequently, the pink curve encloses the largest communication-performance region. The red curve shifts left relative to the green, and the remaining curves are ordered to reflect progressively better communication performance as sensing clutter decreases, indicating that the system tends to apply more exploitation of the available degrees of freedom on communication performance.

\begin{figure}[ht]
    \centering
    \includegraphics[width=\linewidth]{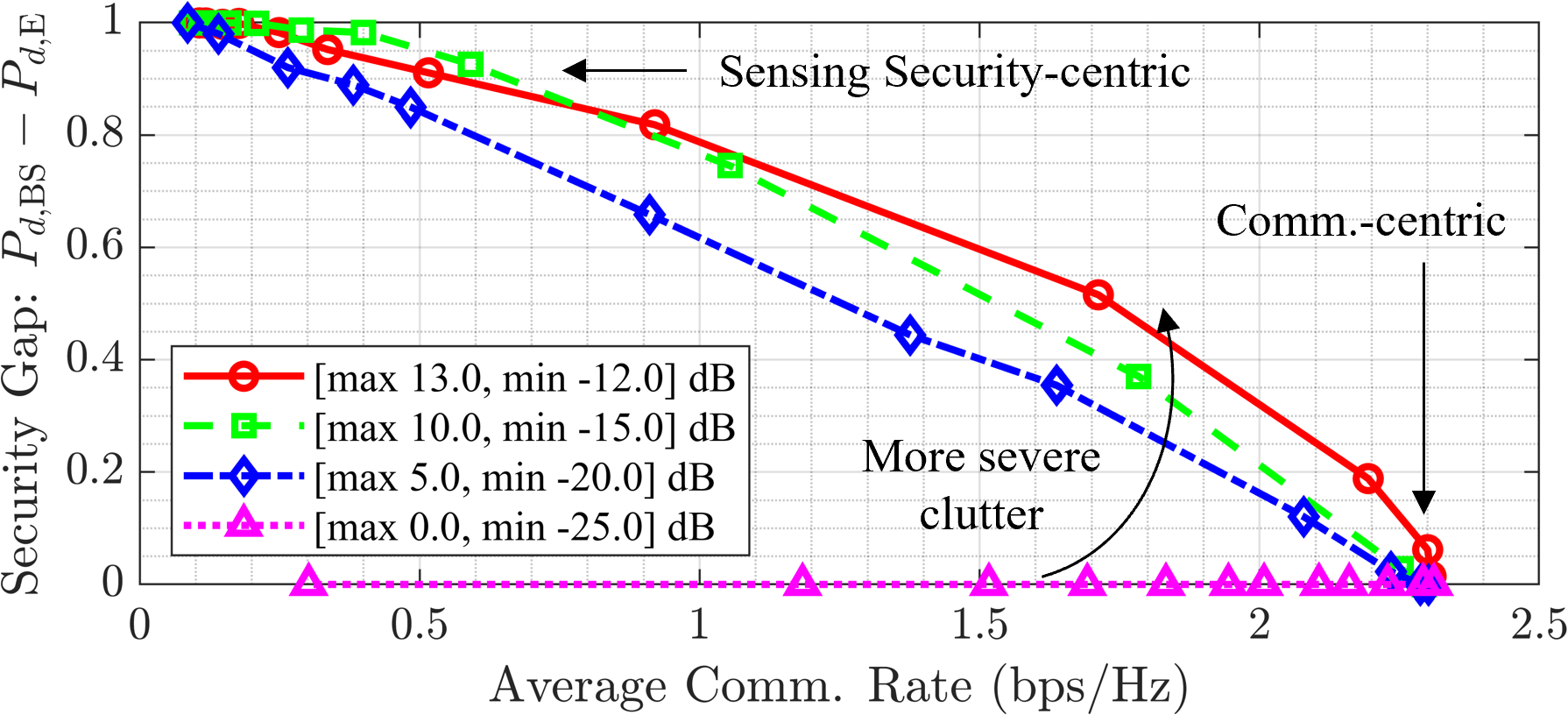}
    \caption{The trade-off for bottom-end metrics of sensing-security and communication under different scatter reflection conditions for OTFS.}
    \label{fig:Trade_off_region}
\end{figure}

Fig.~\ref{fig:Trade_off_region} illustrates the trade-off between bottom-end metrics, specifically the detection probability difference between the BS and Eve, and the average communication rate. As clutter becomes increasingly severe, the performance region initially expands and then, beyond a bottleneck, saturates (see the green and red curves). Thereafter, the boundary reshapes to emphasize communication performance, as extreme clutter compels the optimization to allocate most of the available degrees of freedom to communication, thereby avoiding futile efforts in achieving sensing-security.

\subsection{Comparison Between Eve-Aware and Eve-Agnostic Cases}
\begin{figure}[ht]
    \centering
    \includegraphics[width=0.9\linewidth]{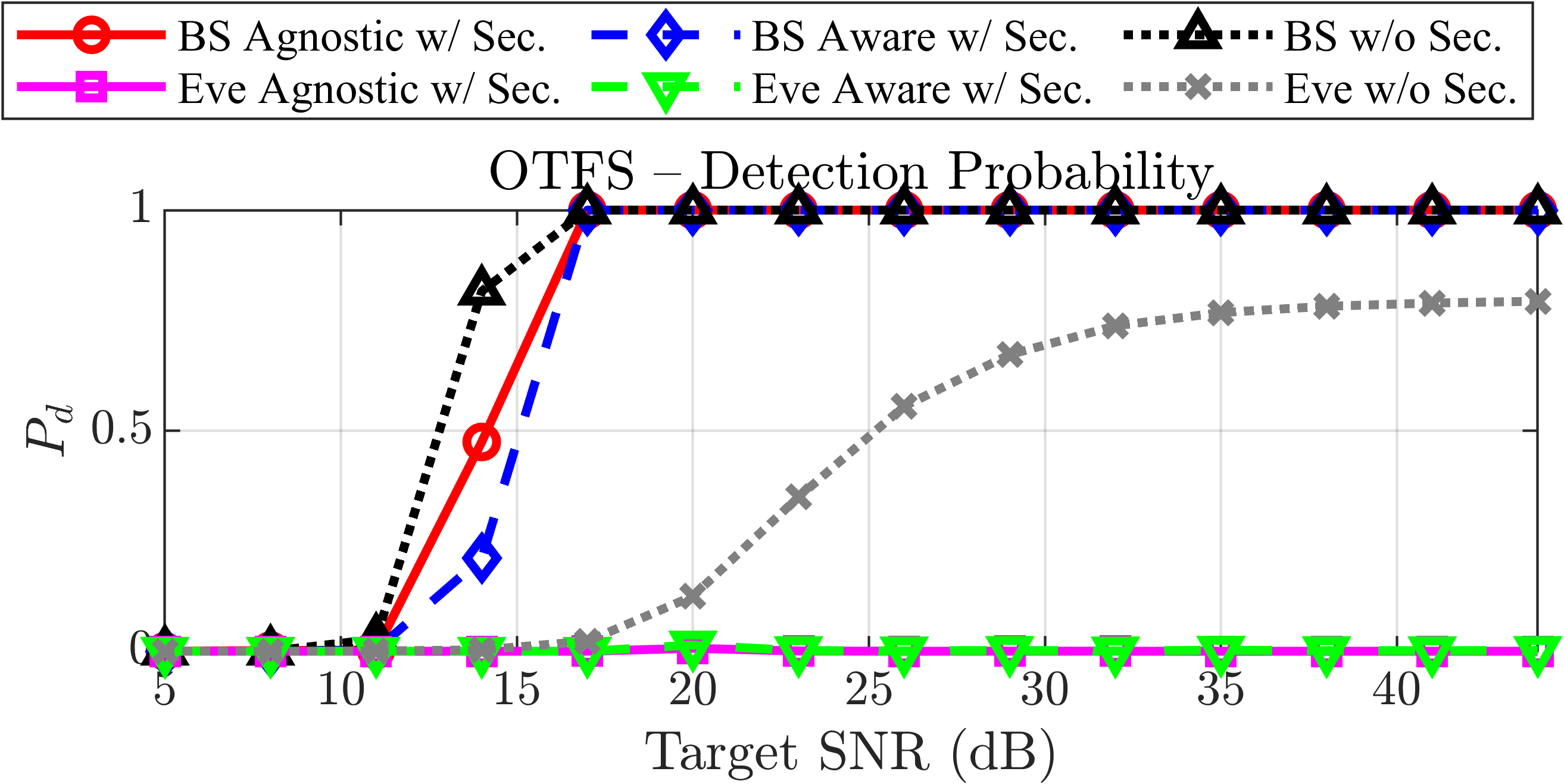}
    \caption{The comparison between Eve-aware and Eve-agnostic cases for the OTFS waveform.}
    \label{fig:EEA}
\end{figure}
Fig.~\ref{fig:EEA} illustrates the applicability of the proposed method to both Eve-aware and Eve-agnostic scenarios, using the same experimental setup as in Subsection~\ref{sec:SS_SNR}. In the Eve-agnostic case, the Eve-related parameters are assigned arbitrary values rather than Eve’s exact parameters. The results show that the proposed method remains effective in this setting, exhibiting only a slight performance difference at the BS compared with the Eve-aware design.

\section{Conclusion}
\label{sec:conclu}
This paper investigated sensing-security for practical ISAC waveforms from a detection-oriented perspective. We established system models for UE communication, BS sensing, and Eve’s passive sensing, and analytically decomposed the RD-domain sensing response into mainlobe, sidelobe, and noise components for both BS and Eve. Building on this, we introduced KLD-based detection metrics and derived a tractable Jensen surrogate to facilitate the analysis and optimization. Leveraging this metric together with a communication-error penalty, we formulated a sensing-security optimization over TF-domain perturbations and solved the resulting nonconvex design via simulated annealing. Numerical results verified that the proposed method reliably suppresses Eve’s sensing by elevating their sidelobes and reducing their detection probability, while causing only a modest performance loss at the BS. The experiments further revealed a clear sensing-security–communication trade-off: stronger sensing-security lowers Eve’s capability at the cost of reduced rate and increased communication error. Overall, the proposed framework provided an effective and practical approach for enhancing sensing-security in ISAC systems.

% if have a single appendix:
%\appendix[Proof of the Zonklar Equations]
% or
%\appendix  % for no appendix heading
% do not use \section anymore after \appendix, only \section*
% is possibly needed

% use appendices with more than one appendix
% then use \section to start each appendix
% you must declare a \section before using any
% \subsection or using \label (\appendices by itself
% starts a section numbered zero.)
%

\appendices
\section{The Approximation for the KLD}
\label{app:KLD_approx}
Under hypotheses \(\mathcal{H}_{0}\) and \(\mathcal{H}_{1}\), the KLD at the BS is
\begin{align}
D_{\text{BS}}
&=\mathbb{E}_{T\sim f_1}\!
   \Bigl[
     \ln I_0\!\bigl(
       \tfrac{2|\mu_{\text{BS}}|\sqrt{T}}
             {\sigma^2_{\text{BS}}}
     \bigr)
   \Bigr]
  -\frac{|\mu_{\text{BS}}|^2}
        {\sigma^2_{\text{BS}}}.
\end{align}
Define the random variable
\[
X = \frac{2|\mu_{\text{BS}}|\sqrt{T_{\text{BS}}}}{\sigma^2_{\text{BS}}},
\]
so that \(\mathbb{E}_{f_{1}}[\ln I_{0}(X)]\) appears in the final term.  
Since \(\ln I_{0}(x)\) is convex for \(x\ge 0\), Jensen’s inequality gives
\begin{align}
\mathbb{E}_{f_{1}}\!\bigl[\ln I_{0}(X)\bigr]
&\ge\ln\! \bigl(I_{0}(\mathbb{E}_{f_{1}}[X])\bigr),\\
\mathbb{E}_{f_{1}}[X]
&=\frac{2|\mu_{\text{BS}}|}{\sigma^2_{\text{BS}}}
  \,\mathbb{E}_{f_{1}}\!\bigl[\sqrt{T_{\text{BS}}}\bigr].
\end{align}
Because the logarithmic derivative of the zeroth‐order modified Bessel function satisfies
\begin{align}
\frac{\text d}{\text dX}\,\ln I_{0}(X)
      &=\frac{I_{1}(X)}{I_{0}(X)},\label{eq:logI0_deriv}\\[2pt]
0 <\frac{I_{1}(X)}{I_{0}(X)}&<1,\quad
\lim_{X\to\infty}\frac{I_{1}(X)}{I_{0}(X)}=1.\label{eq:ratio_limit}
\end{align}
It follows that \(\ln I_{0}(X)\) is nearly affine. Exploiting this property, the KLD admits the following tight Jensen lower bound
\begin{align}
D_{\text{BS}}
    &\approx \widetilde{D}_{\text{BS}},\label{eq:KLD_approx}\\[3pt]
\widetilde{D}_{\text{BS}}
    &=\ln\!\Bigl[
      I_{0}\!\Bigl(
        \frac{2|\mu_{\text{BS}}|}{\sigma^{2}_{\text{BS}}}
        \,\mathbb{E}_{f_{1}}\!\bigl[\sqrt{T_{\text{BS}}}\bigr]
      \Bigr)\Bigr] -\frac{|\mu_{\text{BS}}|^{2}}{\sigma^{2}_{\text{BS}}}.
\label{eq:BS_KL_Jensen_LB}
\end{align}
Let \(R=\sqrt{T_{\text{BS}}}=|z|\) with  
\(z\sim\mathcal{CN}\bigl(\mu_{\text{BS}},\sigma^2_{\text{BS}}\bigr)\).  
Then \(R\) is Rice‐distributed with LoS amplitude \(\nu=|\mu_{\text{BS}}|\) and 
scale \(s=\sigma_{\text{BS}}\).  
Its first moment is
\begin{align}
\mathbb{E}_{f_{1}}[R]
&=\sqrt{\sigma^2_{\text{BS}}}\frac{\sqrt{\pi}}{2}\,
  \exp\!\Bigl(-\frac{|\mu_{\text{BS}}|^{2}}{2\sigma^2_{\text{BS}}}\Bigr)
\notag\\
&\quad\times
  \Bigl[
    \Bigl(1+\frac{|\mu_{\text{BS}}|^{2}}{\sigma^2_{\text{BS}}}\Bigr)
    I_{0}\!\Bigl(\frac{|\mu_{\text{BS}}|^{2}}{2\sigma^2_{\text{BS}}}\Bigr)
    +\frac{|\mu_{\text{BS}}|^{2}}{\sigma^2_{\text{BS}}}
    I_{1}\!\Bigl(\frac{|\mu_{\text{BS}}|^{2}}{2\sigma^2_{\text{BS}}}\Bigr)
  \Bigr].
\end{align}
Substituting the above moment into Eq.~\eqref{eq:BS_KL_Jensen_LB} removes the remaining expectation and yields a fully closed‐form approximation. The KLD for Eve has an analogous pair of bounds obtained by replacing BS parameters with the corresponding Eve parameters.

% you can choose not to have a title for an appendix
% if you want by leaving the argument blank
% \section{}
% Appendix two text goes here.

% % use section* for acknowledgment
% \section*{Acknowledgment}

% The authors would like to thank...

% % Can use something like this to put references on a page
% % by themselves when using endfloat and the captionsoff option.
% \ifCLASSOPTIONcaptionsoff
%   \newpage
% \fi

% trigger a \newpage just before the given reference
% number - used to balance the columns on the last page
% adjust value as needed - may need to be readjusted if
% the document is modified later
%\IEEEtriggeratref{8}
% The "triggered" command can be changed if desired:
%\IEEEtriggercmd{\enlargethispage{-5in}}

% references section

\bibliographystyle{IEEEtran}
\bibliography{bibtex/bib/IEEEabrv,bibtex/bib/IEEEexample}

\end{document}